\documentclass[10pt,journal,compsoc,pdf]{IEEEtran}
\pdfoutput=1

\ifCLASSOPTIONcompsoc
  \usepackage[nocompress]{cite}
\else
  \usepackage{cite}
\fi
\ifCLASSINFOpdf
   \usepackage[pdftex]{graphicx}
   \DeclareGraphicsExtensions{.pdf}
\else
   \usepackage[dvips]{graphicx}
\fi
\usepackage[cmex10]{amsmath}
\usepackage{algorithmic}
  \usepackage[caption=false,font=footnotesize,labelfont=sf,textfont=sf]{subfig}
\DeclareMathOperator*{\argmax}{arg\,max}

\begin{document}
%
\title{Distributed and Fair Beaconing Rate Adaptation for Congestion Control in  Vehicular Networks}
%
%
%
%

\author{Esteban~Egea-Lopez
        and~Pablo~Pavon-Mari\~no
\IEEEcompsocitemizethanks{\IEEEcompsocthanksitem E. Egea-Lopez and P. Pavon-Mari\~no  are with the Department
of Information and Communications Technologies, Universidad Polit\'ecnica de Cartagena (UPCT), Spain,\protect\\
E-mail:$\lbrace$esteban.egea, pablo.pavon$\rbrace$@upct.es}
}

\IEEEtitleabstractindextext{%
\begin{abstract}
Cooperative inter-vehicular applications rely on the exchange of broadcast single-hop status messages among vehicles, called beacons. 
The aggregated load on the wireless channel due to periodic beacons can prevent the transmission of other types of messages, what is called channel congestion due to beaconing activity. 
In this paper we approach the problem of controlling the beaconing rate on each vehicle by  modeling it %
as a Network Utility Maximization (NUM) problem. This allows us to formally apply %
the notion of fairness of a beaconing rate allocation in vehicular networks and to control the trade-off between efficiency and fairness.  
The NUM methodology provides a rigorous framework to  design a broad family of simple and decentralized algorithms, 
with proved convergence guarantees to a \textit{fair} allocation solution. In this context, we focus exclusively in beaconing rate control and  propose the  
Fair Adaptive Beaconing Rate for Intervehicular Communications (FABRIC) algorithm, which uses a particular scaled gradient projection algorithm to solve the dual of the NUM problem. The desired fairness notion in the allocation 
 can be established with an algorithm parameter. Simulation results validate our approach and show that  FABRIC converges to fair rate allocations in multi-hop and dynamic scenarios.

\end{abstract}

}

\maketitle

\IEEEdisplaynontitleabstractindextext

%
\IEEEpeerreviewmaketitle

\IEEEraisesectionheading{\section{Introduction}\label{sec:introduction}}

%
%
%
%
\IEEEPARstart{I}{nter-vehicle} communications based on wireless technologies pave the way for innovative applications in traffic safety,
driver-assistance, traffic control and other advanced services which will make up future Intelligent Transportation Systems
(ITS) \cite{VanetBook}. Communications for  Vehicular Ad-Hoc Networks (VANET) have been  developed and standardized in the last years. At the moment,  a  dedicated short range communication (DSRC) bandwidth  has been allocated to vehicular communications at 5.9 GHz and both American and European standards \cite{EtsiPhy} have adopted IEEE 802.11p  \cite{802.11p} as physical and medium access control layers, based on carrier-sense multiple access with collision avoidance (CSMA/CA). These networks are characterized by a highly dynamic environment where short-life connections between vehicles are expected as well as adverse propagation conditions leading to severe or moderate fading effects \cite{Survey}.  

Cooperative inter-vehicular applications usually rely on the exchange of broadcast single-hop status messages among vehicles on a single control channel, which provide detailed information  about vehicles position, speed, heading, acceleration, curvature and other data of interest \cite{EtsiCAM}. These messages are called \textit{beacons} and are transmitted periodically, at a fixed or variable \textit{beaconing rate}. 
Beacons provide very rich information about the vehicular environment and so are relatively long messages, between 250 and 800 bytes, even more if security-related overhead is added \cite{Awareness}. In addition, vehicles 
exchange other messages on the control channel:  \textit{service announcements} and \textit{event-driven messages} as a result of certain events. For instance, \textit{emergency} messages are transmitted only when a dangerous situation is detected.

The aggregated load on the wireless channel due to periodic beacons can rise to a point where it can limit or prevent the transmission of  other types of messages, what is called \textit{channel congestion due to beaconing activity}. Control schemes are required to prevent this situation and several alternatives are available:  adapting either the beaconing rate,  the  transmission range, the  transmission data rate, the carrier sense threshold or a combination of some of them \cite{Awareness}. In this paper we focus on the control of the beaconing rate of each vehicle.
The goal is to limit the channel bandwidth used by beacons to ensure that the remaining capacity is available to event-based messages.
The practical implementation of the system impose  two strong requirements on the control scheme, in addition to keep the channel load under a desired level: first, to be \textit{distributed} and, second, 
to grant beaconing rates to each vehicle in a \textit{fair} way. Being distributed means that vehicles should control their rate making use only of the signaling information exchanged with their neighbor vehicles and without relying on any centralized infrastructure. Besides, to reduce the signal overhead, the exchanged information should be kept to a minimum.
The control scheme should also provide quick and effective adaptation to changes in the environment, such as  the channel conditions  and  the number of vehicles in range. The limits on such capabilities are captured by the \textit{convergence properties} of the algorithm in use.

\textit{Fairness} must be guaranteed as a safety requirement since  beacons are used to provide vehicles with an accurate estimate of the state of their neighbors \cite{Awareness}. 
Consequently, the fairness goal implies that no vehicle should be allocated arbitrarily less resources than its neighbors, under the constraints imposed by the available capacity. 
However, even starting from the previous principle, 
  several notions of fairness can be defined, and in most of them there is a trade-off between fairness and efficiency  \cite{ChiangFair}: more fairness results usually in a less efficient use of the shared resource. 
But less efficiency is detrimental to safety, since  in general, the higher the beaconing rate, the higher the quality of the state information \cite{Awareness}. 
Thus,  using an inadequate notion of fairness implies not simply wasting resources but also has a negative influence on the safety of the users.
In conclusion, in vehicular networks it is necessary not only to provide fairness but also  to be able to select the \textit{appropriate fairness notion}. 
This ability is a distinguishing feature of the algorithm  put forward in this paper with respect to other proposals \cite{Limeric, Pulsar, Fallah,   Tielert, Intern}.

Several beaconing rate control schemes have been proposed in the literature \cite{Limeric, Pulsar, Fallah}. Although most of them are able to bring the channel load to the desired level, none of them is able to meet all the aforementioned requirements. In particular, all of them consider a very basic approach to fair allocation  of beaconing rates,  
without  a formal  definition and  rigorous convergence support. In some cases, the combination of a basic notion and its particular implementation may result in an unnecessarily low rate and so a decrease in safety as 
we discuss later with an example.
In summary, the issue of fairness is not completely addressed yet. There are two questions related to this issue.  One is which is the appropriate notion of fairness in vehicular networks and 
whether different scenarios require different notions of fairness. Another one is how to enforce a particular fairness notion and control it dynamically. In this paper we focus on the latter one, which has already been 
considered in other contexts as we discuss next. 

Distributed rate control has been extensively studied in other contexts. In particular, \textit{Network Utility Maximization (NUM)}  has received much attention in the field of congestion control in packet switched data networks since the seminal work of Kelly \cite{Kelly99}, and the connection found by Mo and Walrand \cite{MoWal} with fairness in bandwidth allocation. Surprisingly and in spite of the similarities, such an 
 approach has not been adopted for congestion control in vehicular networks. 
Therefore, in this paper we describe a new approach to the problem of beaconing rate control in vehicular networks, modeling it as a NUM rate allocation problem, where each vehicle is associated a so-called \textit{utility function}, 
such that the problem objective becomes the maximization of the sum of utilities of each vehicle. Applying the NUM theory allows us to design a broad family of decentralized and  simple algorithms, with proved convergence guarantees to a fair allocation solution, supported by the rigorous developments of NUM theory. In addition, thanks to the work in \cite{MoWal}, the notion of fairness of a beacon rate allocation in vehicular networks can be formally defined and generalized. The particular (concave) shape of the utility function of the vehicles is related to the different notions of fairness induced globally, the so-called $(\alpha,\omega)$-fairness allocations. As a result, different control schemes can be designed in order to enforce a particular type of fairness, such as proportional fairness ($\alpha = 1$), or max-min fairness ($\alpha \rightarrow \infty$). 
As we will show, NUM modeling allows us to design beaconing rate control algorithms with all the discussed requirements: they are distributed, they require the exchange of a small amount of control information and they can be 
configured to obtain different notions of global fairness, well-defined and with guaranteed convergence properties.
 We propose a particular algorithm 
and validate it with extensive simulations in static and dynamic scenarios. Its performance is evaluated and compared to one 
of the alternative state-of-the-art proposals.

In the remainder of this paper we first review related works in section \ref{Related}. In section \ref{Background} we provide a background on the classical NUM approach for rate allocation in packet switched networks and its connection with fairness. Afterwards, the  beaconing congestion control problem for vehicular networks is formulated as a NUM rate allocation problem in section \ref{Model}, where  we also propose a particular algorithm. In section \ref{Validate},
 it is validated and compared with other proposals in static scenarios.  In section \ref{Results} we extend the comparison and evaluation in different dynamic scenarios. Finally, conclusions and future work are discussed in section \ref{Conclusions}.

\section{Related Work}
\label{Related}

Transmissions in vehicular networks are broadcast in nature and use a CSMA-based medium access control (MAC) with constant contention window and no acknowledgment or retransmission. ETSI standards define a 10 MHz control channel for vehicular communications at 5.9 GHz \cite{EtsiPhy}. Periodic beaconing over one-hop broadcast communications supports cooperative inter-vehicular applications by disseminating status and environmental information to vehicles on the control channel \cite{EtsiCAM}.  
The rate of beacons has an influence on the quality of service of the applications. In fact, some applications may require a certain \textit{beaconing reception frequency}, which is dependent on 
propagation losses, the number of contending  nodes and other considerations, although standards \cite{EtsiCAM} specify the required \textit{beaconing generation rate}. 
A framework for decentralized congestion control (DCC) in the control 
channel has been published by ETSI \cite{EtsiDCC}, which can accommodate a variety of controls such as transmit power, message rate or receiver sensitivity, though the currently  suggested mechanisms are very basic, and extensions are being discussed. 

Regarding pure beaconing rate control proposals,  \cite{Limeric, Pulsar}  propose rate control algorithms that comply with a global generic beaconing rate goal.
 The former, called LIMERIC, uses a linear control based on continuous feedback (beaconing rate in use) from the local neighbors,  whereas the latter, called PULSAR, uses an additive increase multiplicative decrease (AIMD) iteration with binary feedback (congested or not) from one and two-hop neighbors. Both of them, however, show  limitations. 
Regarding fairness none of them define this formally: PULSAR claims targeting  a ``best-effort'' approach to max-min fairness allocation by the use of AIMD,
whereas  LIMERIC  aims to ``achieve fairness such that all the nodes converge to the same message rate''. 
LIMERIC is shown to converge to a  unique equal rate for every vehicle, which is below the optimal proportional fairness rate by design. In fact, there is a trade-off between the convergence speed and the distance to the optimal value. 
And, in any case,  the convergence is only proved when all the vehicles are in range of each other, not for multihop scenarios. 
Regarding PULSAR, it requires synchronized updates and  piggybacking congestion information from vehicles at a two hops.
Authors of LIMERIC  propose to combine the LIMERIC rate adaptation mechanism with the PULSAR piggybacking of two-hop congestion information to achieve global fairness \cite{Limeric}, but it is not proved neither discussed in detail. 
Finally, a recent work \cite{Kim} also shows that both LIMERIC and PULSAR separately  actually may fall into unfair configurations. The authors of \cite{Kim} propose as a solution  heuristic techniques that  ensure that two neighbor vehicles cannot diverge in their allocated rate. As we 
show later, it may actually prevent the algorithm to achieve an optimal fair allocation in some scenarios. 
In summary, these and other techniques proposed to date do not ensure correct convergence to a well-defined fair configuration and have no theoretical support for global convergence.

Transmit power control (TPC) has been investigated in 
recent proposals \cite{SBCC,  Awareness}, which show that TPC can be prone to instabilities, and its accuracy relies on the quality of the propagation model.  
There is a related concept called awareness control: the techniques to adapt the communications
parameters, such as transmit power or beaconing rate, to the requirements of an application, whose goal may be safety or any other such as tracking.
Joint transmit power and rate control
to enforce particular application  quality of service requirements has been studied in \cite{Awareness, Fallah,  EMBARC, Intern}. 
In \cite{Fallah} 
the beaconing rate is set by the requirements of a particular tracking application and authors propose to control
the transmission range in order to keep the channel occupancy within the desired levels.
EMBARC  \cite{EMBARC} is a combination of LIMERIC with a dynamic tracking error control, letting vehicles with higher dynamics transmit with higher rates, while using LIMERIC to adjust the remaining 
capacity to the desired goal. INTERN \cite{Intern} lets the application set the minimum power and rate which requires and uses LIMERIC to equally share the excess of capacity.
Let us note that these proposals do not actually attempt to jointly control both transmission variables, but let the application on top set the minima required and then control the excess of capacity by adjusting  one 
of them. Our proposal is also actually compatible with this approach. Each vehicle can dynamically set its transmit power according to its requirements and each vehicle can independently set a minimum beaconing rate 
as required by an application and, since it is included in the constraints of the optimization problem, it is enforced by the algorithm. In all the proposals this approach obviously prioritizes 
application requirements over strict congestion control, since violations of the desired level of channel utilization may occur in some scenarios \cite{Intern}.
This may be counterproductive in some cases, since the channel might become congested to the point few beacons can be transmitted, and deserves further study which is out of the scope of this paper.
In \cite{Tielert} authors do discuss joint control of power and rate, but no concrete algorithm is provided and fairness is not considered. 

Finally, differentiated quality of service levels can also be 
provided using  beaconing  rate control: with our approach  
different behaviors  can be assigned   to certain subsets of vehicles by either enforcing  \textit{weighted} fairness by selecting appropriate weights ($\omega$) for each vehicle utility function, 
or even by using totally different utility functions for each vehicle. In both cases  vehicles do not need to know the 
weights or utility functions used by other vehicles. 
This combined with the dynamical setting of minimum and maximum rate parameters can be used to implement \textit{prioritized beaconing allocation and congestion control}. 
A vehicle with special needs, such as a platoon leader,  can simply increase the weight of  its utility function or minimum and maximum rate parameters.
Our algorithm will allocate  it higher rates, while reducing the rates of other vehicles if needed to comply with the congestion constraint and still keeping the fairness of the allocation. 
The  study of these matters is left as future work and we focus here on homogeneous utility functions.

\section{Background}
\label{Background}

In this section we describe the key ideas on the NUM modeling for rate allocation in packet switched networks, that lay the foundations of our work in vehicular networks. For more detailed information we refer to \cite{Kelly99,MoWal}. For a deeper background in convex optimization, problem decomposition and its applications to communication networks we refer to \cite{Boyd, Bertsekas, Chiang}, and references therein. 

\subsection{The NUM problem for rate allocation in packet switched networks}

Let $G(N,E)$ be a packet switched network, being $N$ the set of nodes and $E$ the set of links. Let $D$ be a set of traffic sources. For each traffic source $d$, we denote $r_d$ the unknown bandwidth to be allocated to $d$. We denote as $D(e)$ the subset of demands whose traffic traverses link $e$. The \textit{basic} NUM modeling of the rate allocation problem is:
	\begin{subequations}
	\begin{align}
	\max_{r_d} & \sum_d U_d(r_d) \quad \text{subject to: } \\
	& \sum_{d \in D(e)} r_d \leq u_e \quad \forall e \in E \\
	& r_d \geq 0 \quad \forall d \in D
	\end{align}
	\label{eq_num_1}
	\end{subequations}

The objective function (\ref{eq_num_1}a) maximizes the sum of the utility functions $U_d$ of each source. Constraints (\ref{eq_num_1}b) mean that the sum of the traffic traversing a link $e$, should not exceed link capacity $u_e$. Finally, constraints (\ref{eq_num_1}c) prohibit assigning a negative amount of bandwidth to a source. 

Functions $U_d$ for each demand $d$ are strictly increasing and strictly concave twice-differentiable functions of the rate $r_d$ of that demand. Being $U_d$ an increasing function means that sources always perceive more bandwidth as more useful, 
and are always willing to transmit more traffic if allowed. Being concave means that a sort of \textit{diminishing returns} effect occurs in rate allocation, i.e. increasing the bandwidth of a source from $r$ to $r+1$ means a higher increase in utility, than increasing a unit of bandwidth from $r+1$ to $r+2$. The objective function (\ref{eq_num_1}a) is strictly concave, and problem (\ref{eq_num_1}) is a convex program with a unique optimum solution. 

Several problem decomposition strategies allow to  find decentralized implementations of  gradient-based algorithms with convergence guarantees to solve problem (\ref{eq_num_1}). Interested readers can find a surveyed view in \cite{Sri2014} and references therein. In section \ref{Model}, we will use as starting point of our proposal for vehicular networks the dual decomposition of problem (\ref{eq_num_1}), adapting the technique in \cite{Low}.

\subsection{Connection with fairness}
\label{FairnessSect}
As in every resource allocation problem, the optimum rate allocation in a network should balance two competing efforts: maximizing the total network throughput $\left( \sum_d r_d \right)$  and the  fairness of 
the allocation. 
In this context, \textit{fair} means avoiding those allocations where some demands are granted a high amount of bandwidth while others suffer starvation. 

Capturing the essence of what a fair resource allocation is not an easy task, and fairness has been defined in a number of different ways. One of the most common fairness notions is \textit{max-min fairness}. A rate allocation $r$ is said to be max-min fair if the rate of any demand $d_1$ cannot be increased without decreasing the rate of some other demand $d_2$ which in $r$ received less bandwidth ($r_{d_2} \leq r_{d_1}$). Kelly \cite{Kelly99} proposed the concept of proportional fairness. A vector $r^*$ is proportionally fair if for any other feasible rate allocation $r$, the aggregate of the proportional change of $r$ respect to $r^*$ is negative:
	\begin{align*}
	\sum_d \frac{r_d - r_d^*}{r_d^*} \leq 0, \quad \forall r \text{ feasible}
	\end{align*}
That is, the percentages of increases/decreases respect to any other allocation should sum negative. In \cite{MoWal}, Mo and Walrand 
extended the notion of proportional fairness. Let $w = (w_d, d \in D)$ be a vector of positive weight coefficients, $\alpha \geq 0$. A rate allocation $r^*$ is said to be $(\alpha , w)$-proportionally fair if for any other feasible allocation $r$ it holds that:
	\begin{align}
	\sum_d w_d \frac{r_d - r_d^*}{{r_d^*}^\alpha} \leq 0, \quad \forall r \text{ feasible}
	\label{eq_def_alphafairness}
	\end{align}

The importance of the previous definition of fairness is that, if the following utility functions $U_d$ are used, the optimal solutions of NUM rate allocation problems are also $(w,\alpha)$-fair.
	\begin{align}
	U_d (r_d) = \begin{cases} w_d r_d \quad \text{if $\alpha = 0$} \\
				w_d \log r_d \quad \text{if $\alpha = 1$} \\
				w_d \frac{r_d^{1-\alpha}}{1-\alpha} \quad \text{if $\alpha > 0, \alpha \not = 1$} 
		\end{cases}
	\label{eq_num_4}
	\end{align}

This connection was shown in \cite{MoWal}, for the basic NUM rate allocation problem (\ref{eq_num_1}) and  extends to a much more general class of problems.

The $w_d$ values  can be used  to give more importance to the rates allocated to some demands, that is, to achieve  weighted fairness. If all demands are equal for the system $(w_d = 1, \forall d \in D)$, classical fairness notions are produced for some $\alpha$ values. In particular, 0-proportional fair solutions ($\alpha = 0$) are those which maximize the throughput $\sum_d r_d$. Actually, these solutions can be arbitrarily unfair, granting all the link bandwidth to some demands, and zero to others. If $\alpha = 1$ we have the Kelly notion of proportional fairness. In addition, it can be shown that max-min fairness solutions are obtained when $\alpha \rightarrow \infty$ \cite{MoWal}. 

There is no consensus on which particular value of $\alpha$ is best suited for being ``fair enough'' in a resource allocation context. Actually, this decision is clearly application dependent. Lower values of $\alpha$ tend to produce solutions where the amount of traffic carried $\sum_d r_d$ is higher, but with larger differences between the rates allocated to different demands (more ``unfair''). In its turn, higher $\alpha$ values reduce the difference between demands, commonly at the cost of a lower aggregated throughput \cite{ChiangFair}.
In any case, \textit{control  algorithms can enforce the notion of fairness  by setting the $\alpha$ parameter}.

\section{NUM modeling of the beaconing rate control problem in vehicular networks}
\label{Model}

In the previous section we described the classical NUM approach to congestion control in multihop end-to-end transmissions in wired packet switching networks. 
Although the vehicular context, that is, one hop broadcast transmissions,  may seem very different at first sight, in this section 
we show that it is still a valid approach, describe its application and  discuss similarities and differences.  
\begin{figure}
	\centering
		\includegraphics[width=\columnwidth]{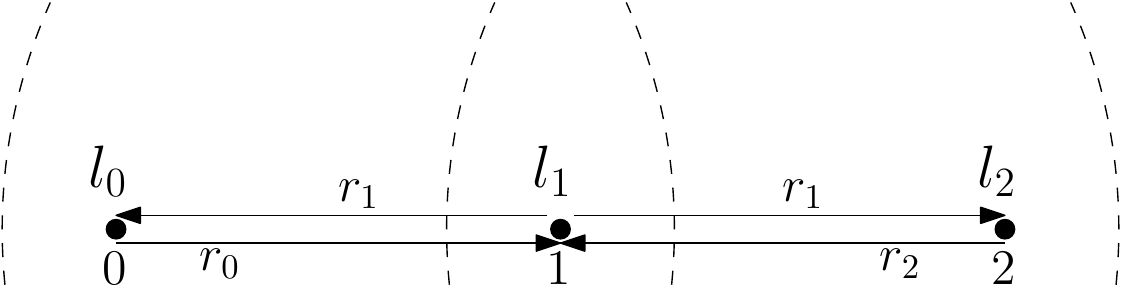}
	\caption{Three vehicles transmitting  $r_0$, $r_1$ and $r_2$ beacons/s respectively. Dashed lines show transmission ranges: vehicle 1 is in transmission range of vehicles 0 and 2, whereas vehicles 1 and 2 are only in 
range of vehicle 1. }
	\label{num-fig}
\end{figure}

As previous considerations,  let us remark that the goal in both cases is to prevent congestion, that is, to keep the usage of resources at a desired level. Now, in our case the resource involved is the wireless channel, which is 
 spatially shared by different nodes, whereas in the classical case the shared resources are network links. Let us also remark that alternative goals may have been defined, such as the maximization of throughput or 
the \textit{reception beaconing rate}, that is, the frequency of correctly received beacons. 
We focus in this paper solely on the 
control of the transmission beaconing rates, which is also the customary approach in other proposals \cite{Limeric, Pulsar}, because: first, to take into account the reception rate we need to 
bring into consideration MAC behavior, which in most practical situations is out of effective control by the user,  and propagation models which require to control the transmit power usually.
 We leave them as future work. 
Second, as pointed out also in \cite{Limeric, Pulsar},  good performance at reception can be indirectly obtained by keeping the  beaconing channel load at a certain level. For instance,  
the information dissemination rate, a performance metric that includes MAC, hidden-node  and physical losses, is maximized when the fraction of channel capacity used is kept at  0.6 to 0.7 \cite{Fallah}.  
A value of 0.6 to 0.7 is also found as the optimum channel load with respect to packet reception in recent works \cite{Subramanian}.

Before formally defining the problem, we provide an informal description of the problem illustrated in Fig. \ref{num-fig}. The key idea is to notice that every vehicle acts as a source, transmitting beacons which use 
a fraction of the wireless channel of their neighbors; and as a resource, defining a wireless channel that is shared with the vehicles in range. That is, we can assign every vehicle a virtual resource, a virtual link $l_v$, 
which has a 
given capacity which is shared by all sources that are in range, including its own  beacons. As can be seen in Fig. \ref{num-fig}, vehicle 1 is transmitting beacons at a rate $r_1$ beacons/s which are 
using the spatially shared channel of both 
neighbors, that is, virtual links $l_0$ and $l_2$ as well as its own virtual link $l_1$. 
Thus, an analogy with packet switching networks would be that the rate $r_1$ is traversing the three links to get to its destination. Conversely  
the total load of resource $l_1$ is the sum of the three rates using it, $r_0+r_1+r_2$. 
So the goal is to assign rates such as the total load of the links they are using is below a certain limit. 
In the following,  we  formally pose the NUM problem for vehicular networks. Let us note that, although useful to get an informal idea,
 we do not use the concept of virtual link for the moment to avoid introducing potentially confusing terms.

Let $V$ be a set of vehicles in a vehicular network. Each vehicle $v \in V$ transmits beacons at a rate $r_v$ beacons/sec, with a constant transmit power. Beacons are broadcast and are received by  neighbor vehicles in 
reception range. Let $n(v)$ denote the set of neighbor vehicles of $v$, which includes $v$. The total rate received by each vehicle is the sum of the rates in its set of  neighbors and we are interested in limiting this amount to a 
maximum $C$, that is, a \textit{Maximum Beaconing Load} (MBL), to avoid channel congestion.

The NUM version of the beaconing rate  allocation problem is given by:
	\begin{subequations}
	\begin{align}
	\max_{r_v} & \sum_v U_v(r_v) \quad \text{subject to: } \\
	& \sum_{v' \in n(v)} r_{v'} \leq C \quad \forall v \in V \\
	& R_v^{min} \leq r_v \leq R_v^{max} \quad \forall v \in V
	\end{align}
	\label{eq_numbeacon_1} 
	\end{subequations}

The objective function (\ref{eq_numbeacon_1}a) is the sum of the utility $U_v(r_v)$ for each vehicle, which depends on the rate $r_v$ allocated to it. 
In order to enforce different fairness notions we take vehicle utility functions as the ones in (\ref{eq_num_4}).
This way we ensure that a rate allocation to the vehicles is $\alpha$-fair if, and only if, it is the optimum solution of (\ref{eq_numbeacon_1}). Constraints (\ref{eq_numbeacon_1}b) mean that the beaconing  load at a given vehicle, which is the one generated by the neighboring vehicles plus its own load ($\sum_{v' \in n(v)} r_{v'}$), must be below the MBL ($C$). 
Finally, constraints (\ref{eq_numbeacon_1}c) force the vehicle rates to be within a minimum ($R_v^{min}$) and maximum ($R_v^{max}$) value. Let us remark that each vehicle can independently set 
its own minimum and maximum rates, which can be used by an application to guarantee a minimum reliability. 

In summary, problem (\ref{eq_numbeacon_1}) reflects  our two goals: (1) to control the congestion while (2)  maximizing the allocated rates in a controllable fair way.
\subsection{Dual decomposition}
\label{DualSect}
In order to find a decentralized algorithm 
solving (\ref{eq_numbeacon_1}) we use  a dual decomposition of the problem. We first form the Lagrangian function $L$ of (\ref{eq_numbeacon_1}) relaxing the constraints (\ref{eq_numbeacon_1}b):
	\begin{align}
	L(r,\pi) & = \sum_v U_v(r_v) + \sum_v \pi_v \left(C - \sum_{v' \in n(v)} r_{v'} \right) = \\\nonumber 
	 &  = \sum_v \left( U_v(r_v) - r_v \sum_{v' \in n(v)} \pi_{v'} \right) +  C \sum_v \pi_v  
	\end{align}

where $\pi_v \geq 0$ are the Lagrange multipliers (prices) associated with the relaxed constraints. 
Multiplier $\pi_v$ is usually interpreted as the price per transmitted traffic unit that other vehicles need to 
pay for occupying the (shared) channel of vehicle $v$, the shared resource. In our particular case, the prices actually reflect the congestion state of the 
link associated to a vehicle, as we discuss later. The Lagrange dual is the maximum value of the Lagrangian over the domain of the rates. 
That is, given a set of non-negative prices $\pi$, the optimal rate allocation solving the Lagrange dual is:
	\begin{align}
	r_v^*(\pi) = \argmax_{R_v^{min} \leq r_v \leq R_v^{max}} \lbrace U_v(r_v) - r_v \sum_{v' \in n(v)} \pi_{v'} \rbrace
	\label{eq_num_21}
	\end{align}

We see that, to compute its rate, each vehicle $v$ needs to know just its own utility function $U_v$ and the set of prices $\pi_{v'}$ of its \textit{neighbor vehicles} and use them to solve problem (\ref{eq_num_21}). Since the original problem is convex with linear constraints, it has the \textit{strong duality property} \cite{Boyd} and the Karush-Kuhn-Tucker (KKT) conditions characterize its optimum solution. Then, it can be shown that there are a set of optimum link prices $\pi^*$ such that the associated rates according to (\ref{eq_num_21}) are the optimal solution of the original problem (\ref{eq_numbeacon_1}). 
In other words, if we obtain the optimum prices, then we can compute the optimal rates with (\ref{eq_num_21}).
The problem of finding such optimum prices is called the dual problem, which can be defined as:
	\begin{align}
	\min_{\pi \geq 0}g(\pi)=\min_{\pi \geq 0} & \lbrace \max_{R_{min} \leq r \leq R_{max}} L(r,\pi) \rbrace
	\label{eq_dualProblem}
	\end{align}

In our case, it can be shown that the objective function in (\ref{eq_dualProblem}), called the dual function,  is strictly convex and differentiable, since the objective function in (\ref{eq_numbeacon_1}) is strictly concave. Thus, the dual problem (\ref{eq_dualProblem}) has a unique set of optimum prices $\pi^*$. 
The classical dual approach for solving the rate allocation problem consists of finding the dual optimal link prices $\pi^*$ using a gradient-based algorithm, as a mean to (in parallel) obtain the optimum rate allocation $r^*$. 

To summarize, in order to find the optimal beaconing rate allocation, vehicles have to exchange their prices $\pi$ with their one-hop neighbors and use them as input to the optimization problem (\ref{eq_num_21}) which can 
be solved autonomously by each vehicle with its local information. 
In  Algorithm 1  we sketch this scheme for vehicular networks, applying a constant-step gradient method,   %
noticing that for the utility functions in (\ref{eq_num_4}) the solution to  (\ref{eq_num_21}) becomes: 
	\begin{align}
	r_v^* (\pi) =\left[ \frac{1}{(\sum_{v' \in n(v)} \pi_{v'})^{\frac{1}{\alpha}}} \right]_{R_v^{min}}^{R_v^{max}} 
	\label{r_eq}
	\end{align}
And that, given a set of prices $\pi$, the gradient of the dual function $g$ evaluated at $\pi$ is given by:
	\begin{align}
	\frac{\partial g}{\partial \pi_v} (\pi) = C - \sum_{v' \in n(v)} r_{v'} (\pi), \quad \forall v 
	\label{subg}
	\end{align}

\vspace{1mm}
\hrule
\vspace{1mm}
\textbf{Algorithm 1}. Beaconing rate control with constant gradient step
\hrule
\begin{algorithmic}[1] %
	\STATE {At $k=0$, set initial vehicle prices $\pi_v^0$ and rates $r_v^0$} 
	\STATE {At each time $k$: } 
	\STATE { Each vehicle $v$ receives the prices of neighbor vehicles $\pi_{v'}^k , v' \in n(v)$. 
		Then: $r_v^k = \left[ \frac{1}{(\sum_{v' \in n(v)} \pi_{v'})^{\frac{1}{\alpha}}} \right]_{R_v^{min}}^{R_v^{max}}$}
	\STATE{ Each vehicle $v$ updates $\pi_v^{k+1}$ according to: } 
\STATE{	$\quad \pi_v^{k+1} = \left[ \pi_v^k - \beta \left( C - \sum_{v' \in n(v)} r^k_{v'} \right)  \right]_0$  }
\end{algorithmic}
\hrule
\vspace{1mm}

We finish this section by  discussing and clarifying relevant aspects of the algorithm and its practical implementation. 
We start with aspects related to the \textit{algorithm and physical meaning of its parameters}.
\begin{itemize}
\item The price ($\pi_v$) reflects the congestion state of the wireless channel of a vehicle and can be thought of as the cost to use the vehicle shared channel. 
Each vehicle measures its own perceived congestion relative to the MBL when the gradient is computed at each iteration $k$ (step 4 of Algorithm 1): The price increases when the channel is congested and 
the other way round.  
\item Note that the implementation of this algorithm is decentralized.  
At each step a vehicle updates  and broadcasts its price $\pi_v^k$ to its neighbors. Then, each vehicle updates its rate using only the information from its one-hop neighboring vehicles. 
It is not necessary to disseminate every price to 
all the  vehicles.
Hence, unlike in packet switching networks \cite{Low}, where the communication of the path prices to the sources may be problematic in a realistic scenario, 
in the case of a vehicular network this solution can be  implemented in practice with little effort 
since the link processing is done by the vehicles themselves, that is each vehicle acts as source and link. 
From another point of view: since transmissions  are mainly single-hop broadcast ones due to periodic beaconing, 
congestion is due to the saturation of the spatially shared wireless channel \textit{only} in the neighborhood of sources (vehicles). 
Therefore, vehicles simply have to broadcast their feedback (prices) to inform all the involved sources of the congestion state of the links they use. 
That is, the sources of congestion of any vehicle are within one-hop distance.  

\item Just by selecting the $\alpha$ parameter different notions of fair allocations are obtained. Moreover,  problem (\ref{eq_numbeacon_1})  can be  used
 to achieve not only different classes of fairness but also to incorporate heterogeneous utility functions and 
constraints for different vehicles. The criteria for selecting a particular fairness 
notion  are application or even scenario dependent. We show and discuss its effect with examples in the following sections but  a deeper discussion
 on the criteria to select $\alpha$ is left as future work. 

\item Convergence of Algorithm 1 can be guaranteed even for asynchronous operation of the vehicles, for a sufficiently small $\beta$, adapting the sufficient convergence conditions in \cite{Low} to our scenario.
To see it, we formally recover the concept of virtual link. Let us assign a virtual link $l \in L$ to each vehicle, $L={1,\dots,V}$, with capacity $C$. Then consider that each virtual link $l_v$ is used by  every vehicle which  
has vehicle $v$ is in reception range with  a rate $r_j$, $j \in n(v)$, plus is own rate $r_v$. That is, we assume that each vehicle virtual link is used by its own rate plus the rates of all the neighbors in range. 
Now, problem (\ref{eq_numbeacon_1})  is equivalent to problem $\mathbf{P}$ in \cite{Low}, where its convergence to the optimal allocation is proved. 

\item The parameter $\beta$ controls the convergence speed of the algorithm, a high  value increases speed but a too high one may cause oscillations. An upper bound can be found in \cite{Low}. Since the algorithm converges any 
initial price $\pi_v^0$ is valid. The values we use for our simulations in next sections have been selected by experimentation. 

\end{itemize}

Next we discuss \textit{implementation and practical aspects}.
\begin{itemize}

\item The procedure is robust against  errors such as packet losses due to fading, collisions or hidden-node interferences. 
We have simulated it  with realistic MAC and propagation models and the results show convergence to the close vicinity of 
the optimal allocation in spite of   packet losses.

\item Each algorithm step $k$ is executed periodically. Vehicles spend a sample period $T_s$ collecting feedback from one hop neighbors and then update their prices and rates according to Algorithm 1. 
The duration of the sample period can be configured and so the absolute convergence time depends on this parameter. We have set $T_s=200$ ms in our simulations.
\end{itemize}

Algorithm 1 can directly solve  problem (\ref{eq_numbeacon_1}),
and its convergence  is guaranteed for a sufficiently small value of $\beta$.
Unfortunately, the theoretical $\beta$ bounds are usually too pessimistic (too small) and in practice much higher values of $\beta$ still result in convergence. 
Since small $\beta$ values mean slower convergence, it is important to have the largest possible step sizes, but it usually results in oscillations. 
Moreover, the gradient update is actually a random process, since gradients are subject to noisy observations and messages can be lost. 
Then, we think it may be advisable to have a conservative approach that do not overreacts to congestion signals, 
and in the next section we  present one possible variation, where we introduce some  modifications and discuss practical considerations.

\subsection{Fair Adaptive Beaconing Rate for Intervehicular Communications (FABRIC)}
\label{FABRICAlg}
In this section, we propose FABRIC (Fair Adaptive Beaconing Rate for Intervehicular Communications), a variation of Algorithm 1, where the prices of the links (Step 2) are updated in a  different form:
	\begin{align}
	\pi_v^{k+1} = \left[ \pi_v^k - \beta sign \left( C - \sum_{v' \in n(v)} r_{v'} \right)  \right]_{0}  
	\end{align}

where $sign(x)$ returns the sign (positive, negative or zero) of the argument.  That is, in this case the vehicle price is increased by a constant  $\beta$  when the channel is congested and decreased by $\beta$ otherwise, 
but never below zero. 
Even though it is not a gradient projection, this variation also converges to the optimal value. In fact, let us note that it can be equivalently considered a scaled gradient 
projection algorithm \cite[Sec. 3.3.3]{Bertsekas}, where the price update is done 
by $\pi_v^{k+1} = \left[ \pi_v^k - \beta M(k)^{-1} \left( C - \sum_{v' \in n(v)} r^k_{v'} \right)  \right]_{0} $, and the sequence of symmetric  positive definite diagonal matrices $M(k)$ is given by 
	\begin{align}
	\left[ M(k) \right]_{ii} = \begin{cases} \left| sg_i(k) \right| & \quad \text{if $\left| sg_i(k) \right| > 0$} \\ 1 & \quad \text{if $\left| sg(k) \right| = 0$} \end{cases}
	\end{align}
 
with $sg_i(k)= C - \sum_{v' \in n(i)} r^k_{i'} $. Therefore, the algorithm meets the conditions given in \cite{Bertsekas2} for convergence to a point arbitrarily close to the optimum, for $\beta$ small enough.
FABRIC  aims to have a reasonably fast convergence compared to that of a standard gradient algorithm (Algorithm 1) while limiting the maximum variation of the rates using the sign function. 
This is a conservative approach, to smooth out wide variations of beaconing rate that may occur because of the noisy observations in which we base the gradient updates.  
In any case, since there are multiple possibilities \cite{Bertsekas}, we think that it should be considered a useful one among other alternatives, whose evaluation we leave as a future work.

Let us now discuss  implementation details. The perceived congestion, that is, the difference between channel capacity and the  fraction used can be obtained in several ways, either by monitoring the channel, that is, measuring the \textit{Channel Busy Time} (CBT), or by counting the number of correctly received beacons. 
In both cases, the result is an estimate of the real channel occupation, since it depends on channel conditions and collisions. Another possibility is to let vehicles inform others of their current beaconing rate by piggybacking it in the beacon. This is our choice and  in our opinion 
the more reliable option with regard to the accuracy of the rate control, since it informs about the actual offered load of the channel in absence of errors, such as fading or interference. 
Moreover, when packets are lost, it provides additional robustness against  noisy measurements. Even though some beacons may get lost, receiving at least one beacon during the sampling period is enough to recover 
the actual price and rate  used by a neighbor.
On the other hand, the algorithm requires  each vehicle $v$ to store and send a non-negative real number, its price $\pi_v$. So, regarding \textit{overhead},  vehicles should broadcast at most their current beaconing rate plus the price, both piggybacked in a beacon, which adds little overhead to the current procedures, for instance, two 32-bit extra fields, around 1\% for  500-byte beacons. To compare, 
LIMERIC combined with PULSAR requires to piggyback in a beacon the locally measured CBT and the maximum CBT reported by the one-hop neighbors, that is, two real numbers, so the overhead is exactly the same as the FABRIC one.

Second, we consider \textit{synchronous or asynchronous} implementations. In the first case, all the vehicles update their rates at the same instant with the received prices. This is possible in practice for vehicles equipped with a GPS device. In that case, all the neighbor prices are available prior to the beaconing rate update. In the second case, each vehicle may update its rate at different instants. Thus, some vehicles may not have all the updated prices from their neighbors and oscillations may occur, which are called flapping. 
To avoid them, we propose to use an  \textit{anti-flapping parameter}, $f$, so that we consider that a gradient coordinate is 0 when its absolute value is below  a fraction $f$ of the capacity. This way vehicles lock their prices when they are close to the MBL, and rate oscillations are prevented. In Section \ref{Validate} we preliminary validate this procedure experimentally, but leave a more detailed study for a future work.

\section{Validation}
\label{Validate}
In this section we test the validity of our algorithm and assumptions, in a static  scenario  where vehicles do not move which allows us an accurate control of the vehicles interactions. The results of FABRIC are compared with 
those of LIMERIC \cite{Limeric}. However, LIMERIC properties have only been proved so far 
for single hop scenarios and, in fact, our simulations show that unfair allocations are obtained in multihop scenarios. Thus, in order to provide a 
fair comparison, we have simulated LIMERIC combined with PULSAR (referred to as LIMERIC+PULSAR when used together) as the authors of LIMERIC suggest \cite{Limeric}, according to the details of \cite{itsworkshop}. 

\textit{Simulations setup}. We summarize first the simulation parameters that are common to the simulations studies in this and the following section. 
The simulations have been implemented with the  OMNET++ framework and its inetmanet-2.2 extension \cite{omnetpp}, which implements the 802.11p standard. This library also implements a realistic propagation and interference model for computing the Signal to Interference-plus-Noise Ratio (SINR) and determining the packet reception probabilities, considering also  capture effect. 

\begin{table}[!t]
	  \caption{Common Parameters for Simulations }
	  \label{tcbt}
\begin{center}
\begin{tabular}{ |c | c | }
\hline                       
Parameter & Value \\
\hline                       
Data rate ($V_t$) & 6 Mbps \\
Sensitivity (S) & -92 dBm \\
Frequency & 5.9 GHz\\
 Power & 251 mW \\
Noise  & -110 dBm \\
SNIR threshold ($T$) & 4 dB \\
Neighbor Table update time & 1 s \\
Sample period $T_s$ & 200 ms \\
Beacon size ($b_s$)  & 500 bytes  \\
Maximum Beaconing rate  ($R_v^{max}$)  & 10 beacons/s \\
Minimum Beaconing rate  ($R_v^{min}$)  & 1 beacons/s \\
$\beta$  &  $2.8\times10^{-5}$\\
$\pi_v^0$  & $1.252\times10^{-3}$ \\
$ f$  & $0.22$ \\
$\alpha_L$ (LIMERIC) & $0.1$ \\ 
$\beta_L $ (LIMERIC)  & $1/150$	 \\ 
\hline  
\end{tabular}
\end{center}
\end{table}

In our tests, vehicles are located on a straight single lane road and their positions are either \textit{deterministic}, that is equally spaced with distance $d$ m or randomly positioned according to a \textit{Poisson} distribution of average density $\rho$ vehicles/m. 
Both \textit{free space} and  \textit{Nakagami-m} propagation models have been used. In both cases, the path loss exponent has been set to  2 or 2.5 depending on the scenario. These are slightly lower values than those reported by \cite{ChanJSAC}, measured in suburban scenarios. Higher values result in shorter transmission range and so congestion is more unlikely and its effects milder. Thus, our values model a worst case scenario. 
Nakagami-m shape parameter has been set to $m=1$ or $m=3$, to model severe (Rayleigh) or moderate fading conditions. 
The MBL has been set to 3.6 Mbps, which is 60\% of the available data rate of 6 Mbps. We  use a beacon size of 500 bytes plus 76 bytes of 
MAC and physical headers,  which results in a maximum channel capacity of $C=781.25$ beacons/s. Table \ref{tcbt} summarizes the rest of common parameters, used unless another value is explicitly mentioned in the text. 
All the simulations  have been replicated 10 times with different seeds. %
\begin{figure}[!t]
	\centering
		\includegraphics[width=\columnwidth]{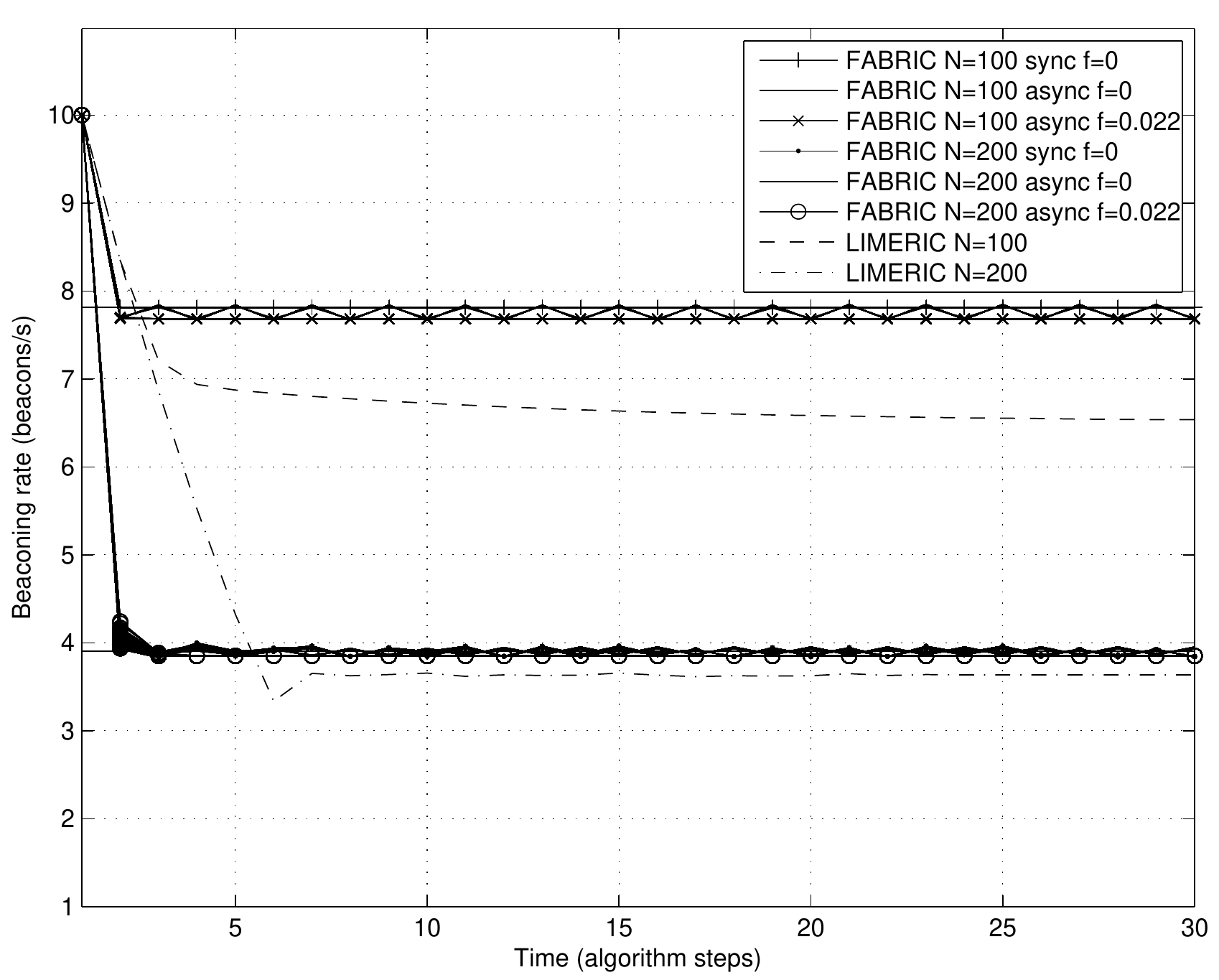}
	\caption{Selected beaconing rate in a one-hop scenario for FABRIC and LIMERIC with 100 and 200 vehicles and with and without anti-flapping 
($f$) parameters. LIMERIC parameters have been set to $\alpha_L=0.01$ and $\beta_L=1/150$. Optimal values shown with a straight line. }
	\label{basic}
\end{figure}

\subsection{ All vehicles  in range of each other}
\label{AllInRange}
In the first scenario vehicles are positioned along a 1000 m long line with a Poisson distribution of average density $\rho=0.1$ and $\rho=0.2$ vehicles/m, with $N=100$ and $N=200$ vehicles respectively. The propagation model is a deterministic free space model, with path loss exponent of 2 and vehicles using 1000 mW of transmit power, which makes all of them to be in range of each other.
The optimum value for the beaconing rate  is $r^*=C/N$, that is $r^*=7.8125$ and $r^*=3.906$ beacons/s respectively. In Fig. \ref{basic} we show the evolution of the beaconing rate with time, in algorithm steps. Vehicles  update their beaconing rate every  $T_s=1$ s in this scenario. In the synchronous case all the vehicles perform their updates at the same time instant, whereas in the asynchronous one the instant is uniformly distributed along the period. As can be seen, FABRIC quickly converges to the optimum value without oscillations in the synchronous case, whereas in the asynchronous some oscillations can occur, that are corrected with the use of the anti-flapping technique described in the previous section. The amplitude of the oscillations decreases with the number of vehicles since the relative weight of the outdated prices is lower in the updates. 

Regarding LIMERIC, Fig. \ref{basic} shows that, although it assigns the same fraction of the bandwidth to all vehicles, such fraction  is 15\% below and 7\% below the optimal one respectively.  It is noticeable that LIMERIC does not achieve the optimal value \cite{Limeric} even in an ideal scenario like this one. The reason is that the LIMERIC operation is controlled by two parameters $\alpha_L$ and $\beta_L$, such that by design the rate allocation uses a fraction of the available channel capacity equal to $\frac{N \beta_L}{\alpha_L + N \beta_L}$. Better utilizations result when $\alpha_L$ is small respect to $\beta_L N$, but this also results in slower convergence times. 
The values used in this paper are the ones suggested by the authors in \cite{Limeric}.

\subsection{ Multihop ideal  scenarios and  differences between fairness notions} 
\label{MultihopSect}
\begin{figure*}[!t]
	\centering
\subfloat[Ideal multi-hop scenario. Top: exact optimal values computed by JOM. Bottom: FABRIC and LIMERIC+PULSAR]{
		\includegraphics[width=0.96\columnwidth]{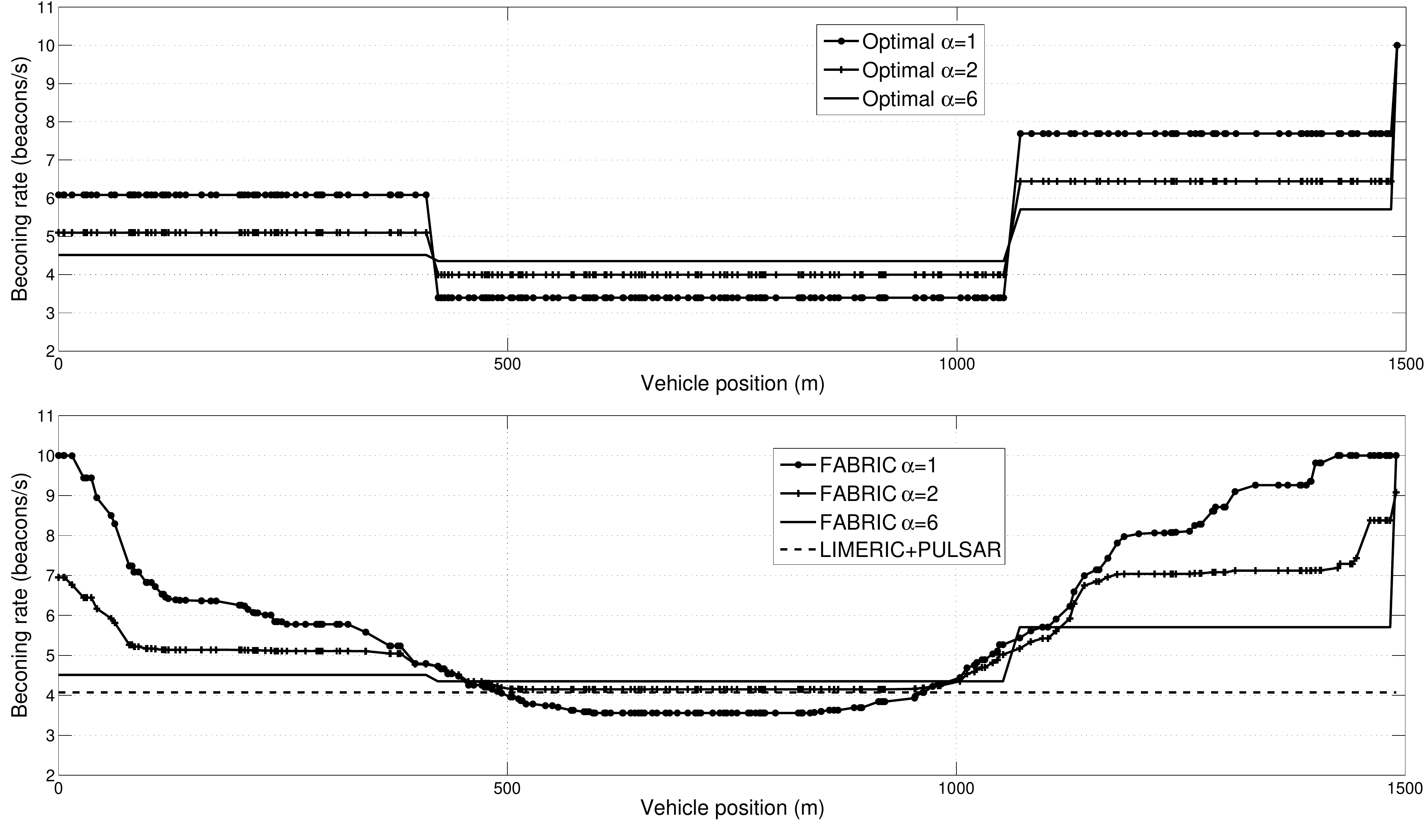}%
	\label{ideal-lin-cbt}
}
\hfill 
\subfloat[Two clusters of vehicles approaching a traffic jam. Cluster B is in range of all the vehicles in Cluster A and a fraction of the vehicles in the traffic jam. A  schematic diagram is shown below.]{
		\includegraphics[width=0.96\columnwidth]{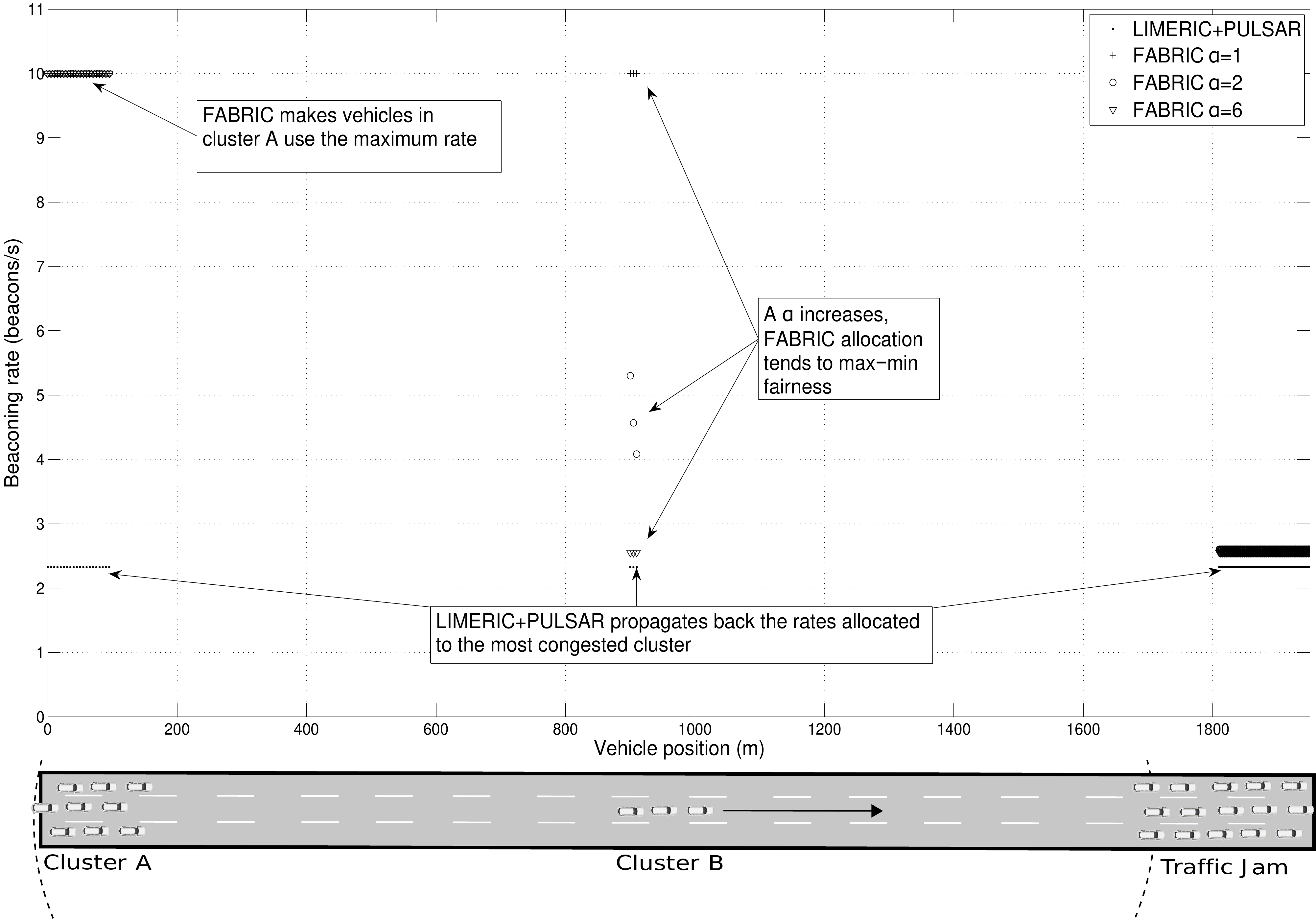}%
	\label{ideal-clust}
}
	\caption{Beaconing rate  versus position in an ideal multi-hop scenario and traffic jam  scenario for  FABRIC and LIMERIC+PULSAR.}
	\label{ideal-multi}
\end{figure*}
%

In these tests, we evaluate two more demanding scenarios where not all vehicles have the same number of neighbors. In addition, we compare and discuss how the fairness parameter $\alpha$ results in different allocations. 
An ideal channel is considered and packets are not lost. 
In the first one, vehicles are  positioned along a 1500 m long line, with a Poisson distribution of average density $\rho=0.14$ vehicles/m. 
In Fig. \ref{ideal-lin-cbt} we plot the results for the  beaconing rate selected versus the position on the road of the vehicles.  

We also plot the exact optimal allocations calculated by solving the NUM problem (\ref{eq_numbeacon_1}) with a numerical optimization solver, provided by JOM \cite{JOM}. 
Proportional fairness is obtained with $\alpha=1$ and as $\alpha$ is increased the allocation tends to max-min fairness. We plot results for $\alpha=2$ and $\alpha=6$.

\textit{Beaconing rates}. Results show that FABRIC  converges to a solution close to the optimal value in all cases, being better as $\alpha$ decreases.
The reason is that when the optimal allocation shows pronounced differences between the rates of neighbor vehicles at some points, as occurs with proportional fairness here, the algorithm needs more steps to converge.  
The solution in Fig. \ref{ideal-lin-cbt} has been achieved after  100 steps (20 s) of the algorithm. 
 Note that an scenario like the one shown, in which the optimal proportional fairness allocation has sharp differences between neighbor rates, is a counter-example to the proposal of \cite{Kim}, 
which suggests enforcing fairness by limiting the differences in the rate allocations of neighbor vehicles.

The convergence to the \textit{exact} optimal allocation may be long, specially if the optimal allocation shows a jagged shape, but 
from a  practical point of view, for randomly positioned scenarios and realistic propagation models, it is not necessary  truly maximize the utility function 
 to obtain acceptable results. That is, even though the allocation is not yet optimal the CBT is already below the MBL and the allocation is close to the optimal,  
for instance, as shown in next sections, for $\alpha=1$ the root-mean-square error between the rates and the optimal allocation is 1.4951 beacons/s after 20 steps (4 s) 
and 80\% of the vehicles measure a CBT below the MBL.

LIMERIC+PULSAR assigns all the vehicles exactly the same rate. In fact, this is the expected behavior in any scenario, since, with PULSAR, the maximum CBT experienced by both one and two-hop neighbors is used as a 
feedback signal by LIMERIC. In the particular scenario of Fig. \ref{ideal-lin-cbt}  this behavior can be consistent with achieving max-min fairness.

\textit{Fairness}.
 As $\alpha$ increases, the optimal allocation tends to assign the same rate to 
all vehicles, trading efficiency by fairness \cite{ChiangFair}. Fig. \ref{ideal-lin-cbt} clearly shows how the allocation becomes flatter as it tends to max-min fairness. 
It also shows that the fairness degree of the allocation can be effectively controlled by FABRIC.


Whether is preferable to set proportional, max-min or any other $\alpha$-fairness is a matter of discussion and possibly application or scenario dependent. For instance, an scenario like this one may model a traffic jam. 
Since vehicles at the edges of the jam may be more exposed to other vehicles approaching at high speed, it  may be desirable to use proportional fairness since it assigns a higher beaconing rate at the border vehicles
In any case, the key advantage of FABRIC is that it can be configured with the $\alpha$ parameter to achieve any of the fairness notions. Moreover,  vehicles can use different fairness goals simultaneously and the parameters can even be dynamically set. 

To emphasize the importance of an adequate selection of the fairness notion in vehicular networks and its influence on the safety of users we provide the 
scenario shown in Fig. \ref{ideal-clust} as a simple but illustrating example.
We have a traffic jam with a high density of vehicles and two clusters of vehicles approaching it, separated by a distance of 900 m
With a transmission range of 1000 mW, the three vehicles of cluster B are in range of both the first 14 vehicles in the jam and of all the 20 vehicles in cluster A. 
There are 150 vehicles in the traffic jam, all in range, resulting in a high congested channel. 

Here the basic approach to fairness of LIMERIC, that is assigning all vehicles an equal rate, combined with the PULSAR implementation which uses the maximum CBT observed within two hops, makes that 
vehicles in cluster A, 2000 m away and with no congestion in the channel, use the same beaconing rate of 2.35 beacons/s that vehicles in the traffic jam. Hence, it unnecessarily reduces the 
beaconing rate, which may affect the safety if those vehicles are driving at high speeds.
On the contrary, with FABRIC, vehicles in cluster A correctly use the maximum beaconing rate of 10 beacons/s, since their channel is not congested at all. Moreover, the 
rates used by cluster B vehicles depend on the choice of $\alpha$, that is, the fairness notion selected. With $\alpha=1$ the maximum rate is used and as it increases, the rate is decreased, being equal to 
the one in the traffic jam (2.608 beacons/s) only when max-min fairness ($\alpha=6$) is selected as  expected, and only for the first vehicle in the cluster.

Except for the distances used in the example, it is not an unlikely scenario which shows the need for mechanisms that provide well-defined fairness control in vehicular networks. 
FABRIC provides this control as a first step but a study on the proper applicability of the fairness alternatives and its dynamical setting  is left as future work.


\subsection{ Realistic scenario with hidden nodes and packet losses}
\label{mhsimscenario}
\begin{figure*}[!t]
	\centering
\subfloat[Allocated beaconing rate  and CBT versus vehicle position. ]{
		\includegraphics[width=0.96\columnwidth]{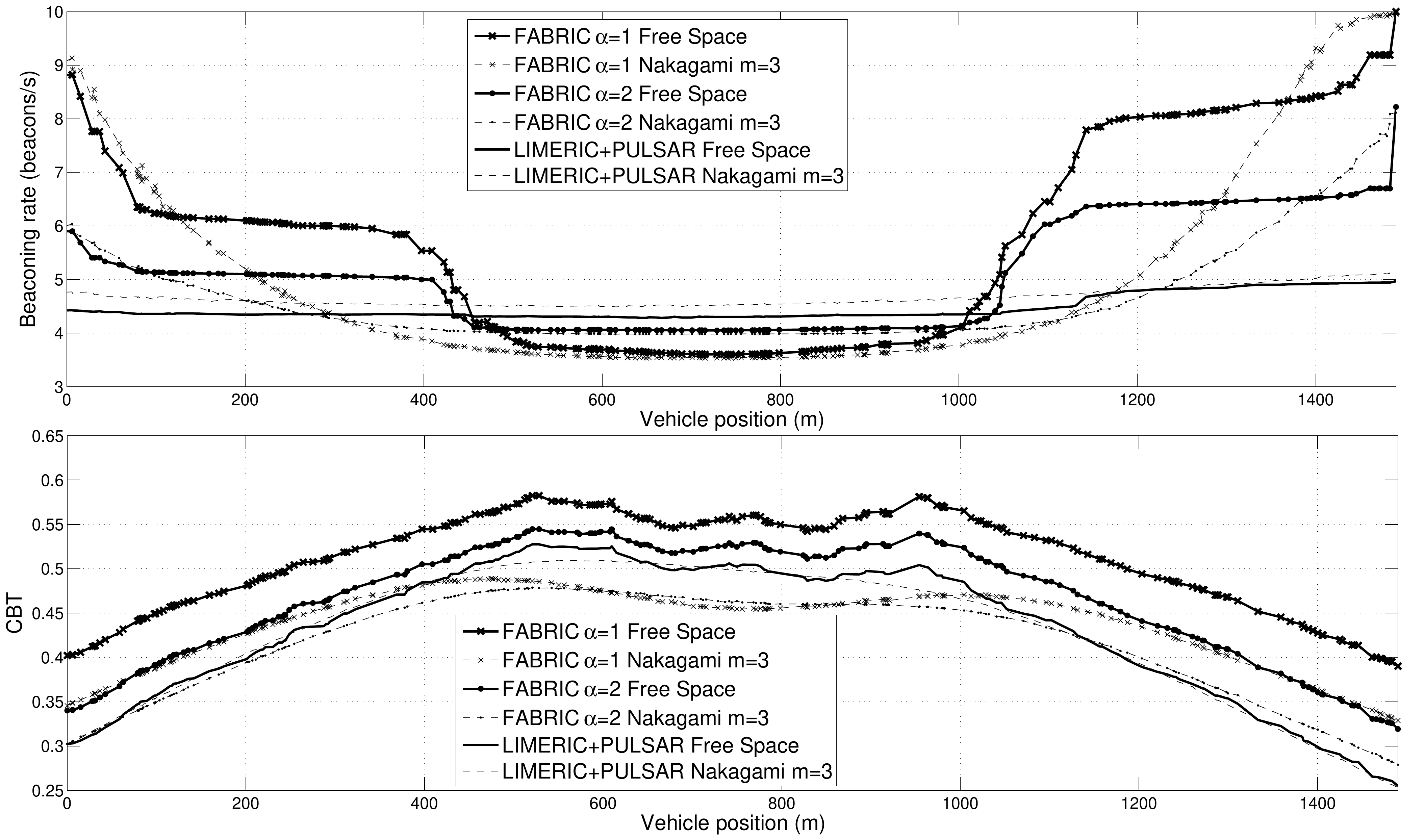}%
	\label{brmulti}
}
\hfill 
\subfloat[Effective beaconing rate  delivered at 250 m,  $\hat{r}_v(250)$, and average IRT versus vehicle position.]{
		\includegraphics[width=0.96\columnwidth]{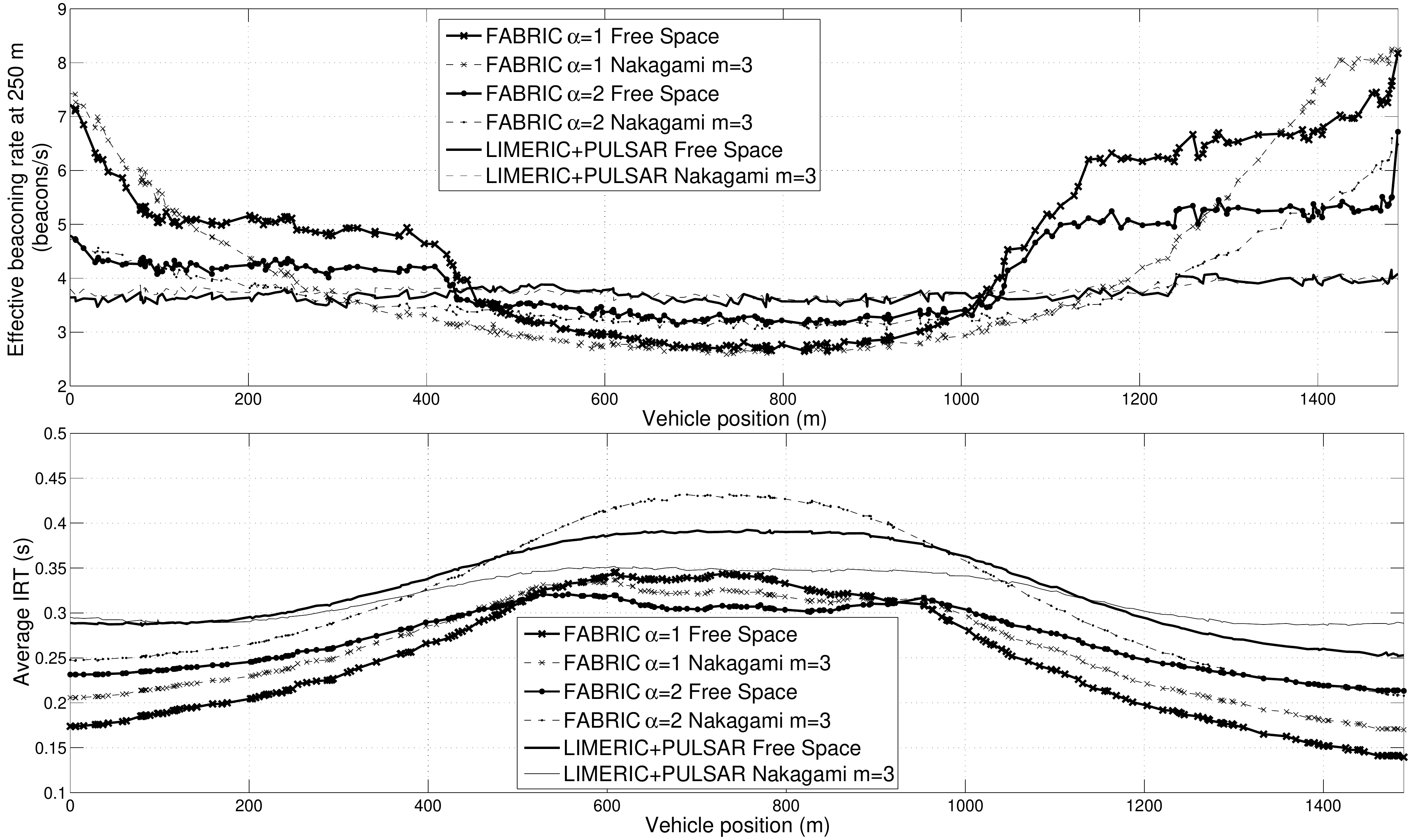}%
	\label{efbrmulti}
}
	\caption{Realistic  multi-hop scenario  at   $t=18$ s, with Free Space and Nakagami m=3, for FABRIC asynchronous   
and  LIMERIC+PULSAR.   }
\label{multihop}

\end{figure*}
In this scenario vehicles  we use the same positioning as in Scenario 2 and test the effects of a realistic environment. The propagation models have been set to free space and Nakagami-m with path loss exponent equal to 2.5. In both cases, the transmit power has been set to 251 mW which results in a transmit range of 531.5 m for free space and an average of 510.5 m for Nakagami-m \cite{SBCC}. 
Let us remark that, with this setting, there are interferences due to hidden nodes. In fact, in this scenario we can see the effects of MAC and hidden collisions as well as fading and channel errors in our model, since it 
has been simulated with an accurate 802.11p and propagation model, including SINR evaluation and capture effect.

\textit{Beaconing rate}. In Fig.  \ref{brmulti}  we show the allocated  beaconing rate  for each algorithm. Confidence intervals at 95\% have been computed but, since the maximum radius of the confidence interval does not exceed 0.05 and 0.3 
beacons/s for free space and Nakagami-m respectively, they are not shown to avoid cluttering the figure.  
 Again, as $\alpha$ increases the optimal allocation tends to become flat. 
With free space propagation, after 18 s  asynchronous FABRIC  approximately tracks the optimal allocations (shown in Fig. \ref{ideal-lin-cbt}, previous section) for $\alpha=1$ and $\alpha=2$ 
and the realistic environment tends to smooth the allocation.
 LIMERIC+PULSAR again removes almost all 
the differences between node rates. In both cases, FABRIC works properly in spite of hidden node collisions and interferences.

\textit{Fading effects}. With fading (Nakagami-m), FABRIC reduces the beaconing rate globally, whereas LIMERIC+PULSAR increases it. This opposite behavior is due to the different feedback signals used. LIMERIC+PULSAR uses directly the measured CBT. 
In this case, fading results in a decrease of the measured CBT and hence LIMERIC increases linearly the rate,  5\% on average consistent with an equal lost of packets due to fading \cite{SBCC}. On 
the contrary, FABRIC uses the  price piggybacked in received beacons. There is now a chance of receiving beacons from far away nodes and all received prices are used to compute the next rate in step 3 of Algorithm 1. 
In fact, we have measured an average increase of 10\%  in the time-averaged number of neighbors for the fading scenario. Consequently, the allocated rate is reduced, particularly on the borders, whereas in the middle nodes,  whose number of neighbors change less, is kept equal.  

\textit{CBT}. The MBL constraint is met for  all vehicles and cases at t=18 s, as shown in Fig. \ref{brmulti}. In fact, we show in next sections that CBT is below the MBL for most of the vehicles 
only after a few seconds. As expected, proportional fairness ($\alpha=1$) provides the more efficient allocation, closer to the maximum allowable use of the channel, at the cost of less fairness. LIMERIC+PULSAR is 
driven by CBT and hence the reduction due to fading (Nakagami-m) is compensated by an increase in rates (Fig. \ref{brmulti}), keeping the CBT at the same level. On the contrary, the effect of fading on FABRIC, as discussed before, 
reducing the rates results in a global reduction of CBT. If necessary,  this behavior  can be corrected in the implementation of FABRIC by filtering  unreliable links: using only the reported prices from
neighbors whose beacons are received a certain number of times, for instance. Testing it has been left as future work.

\textit{Performance}. To better
 measure the effectiveness of the algorithms in a lossy scenario, we  define the \textit{effective delivery ratio at distance $d$}, $D_v(d)=\frac{c_v(d)}{n^S_v(d)} $, where  $c_v(d)$ is 
 the total number of correctly received copies of a beacon up to a distance  $d$ of the transmitter $v$ and $n^S_v(d)$ is the total number of copies of a beacon whose power is above the sensitivity up to that distance. 
That is, $D_v(d)$ indicates how many copies  of a broadcast beacon  are correctly received at a distance no greater than $d$ from the transmitter. 
We then define \textit{effective beaconing rate at distance $d$}, as $\hat{r}_v(d)=r_v \bar{D}_v(d)$, 
that is, a measure of the actually received beaconing rate at a certain distance of the transmitter. Thus,  $\hat{r}_v$ is  a performance metric from the point of view of the transmitter.
From the point of view of the receiver we show the average Inter-Beacon Reception Time (IRT) measured by the vehicles. 

From Fig. \ref{efbrmulti} we see a reduction  around 20\% to 25\% at 250 m, $\bar{D}_v(250)$. The losses are mainly caused by  collisions and 
so the reduction in delivered beacons  is more pronounced for the border nodes,  whose receivers have more potential hidden nodes, and is almost equal for 
all the proposals evaluated. It reflects that the higher rates allocated by FABRIC at the border nodes are balanced by the lower ones at the middle nodes in terms of causing hidden-node collisions.
For comparison, the same scenario with no beaconing control, all vehicles transmitting at 10 beacons/s results in a CBT of 0.9 and 30\% of collided packet in the middle section. 
The IRT for LIMERIC+PULSAR, since all the vehicles use practically the same rate, directly shows the losses due to  collisions with Free Space and a mixture of collisions 
and fading with Nakagami-m. With FABRIC, it includes all those effects and adittionally averages the different rates used for vehicles at different positions. 
FABRIC actually outperforms the measured IRT of  LIMERIC+PULSAR, except for $\alpha=2$ and Nakagami in the midle part of 
the scenario. The allocation of FABRIC with high rates at the borders and low ones in the middle results in fewer collisions and a lower average IRT.
 

%

\section{Dynamic Scenarios}
\label{Results}
\begin{figure*}[!t]
	\centering
\subfloat[Beaconing rate versus time for a single vehicle approaching the cluster of vehicles at 32 m/s. ]{
		\includegraphics[width=0.98\columnwidth]{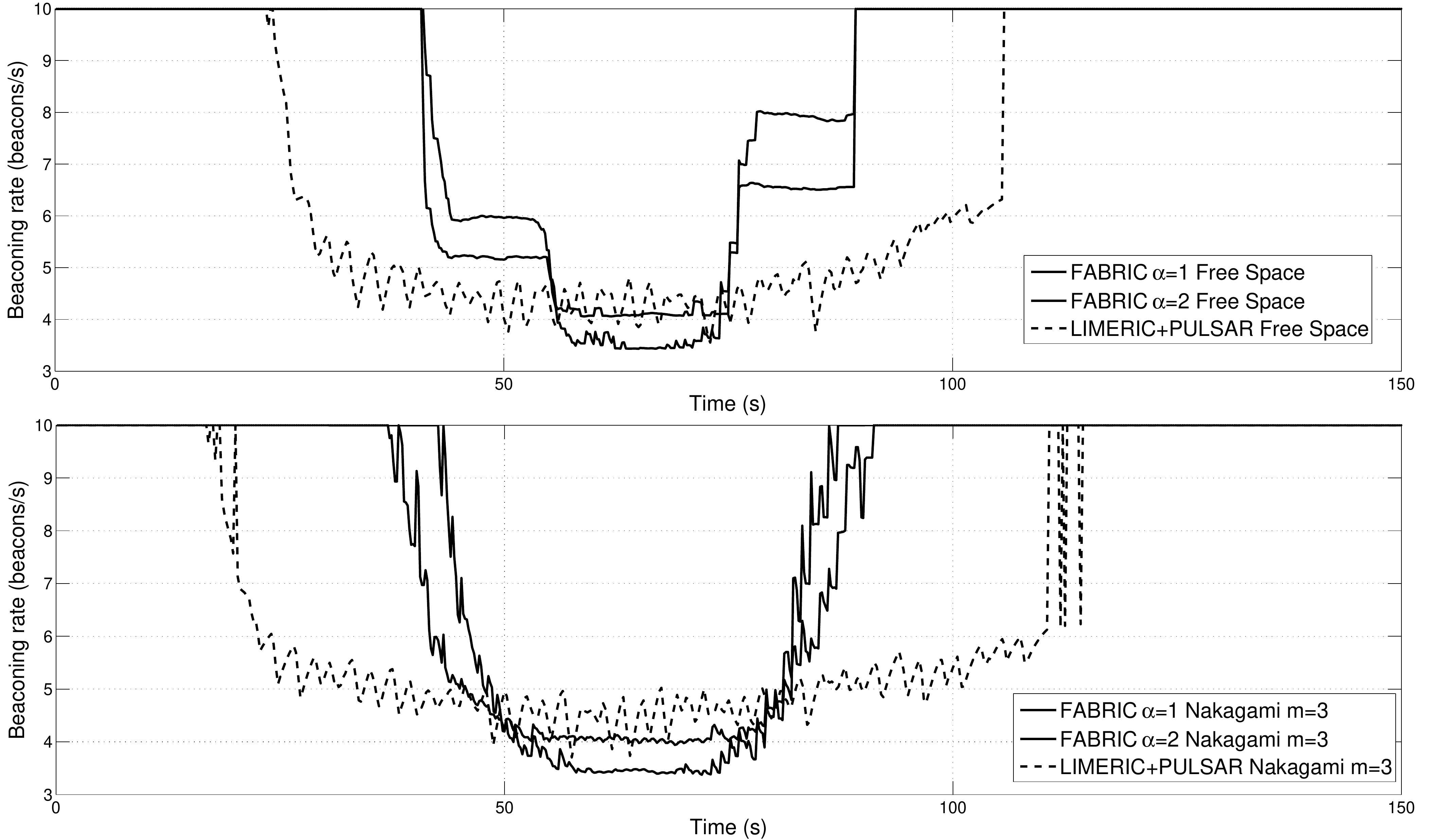}%
	\label{onramp}
}
\hfill 
\subfloat[CBT versus  time and vehicle number in the cluster. FABRIC with $\alpha=1$ and  Free Space propagation. A horizontal plane at z=0.6 has been plotted to mark 
the MBL. A schematic diagram is shown below.  ]{
	\includegraphics[width=0.96\columnwidth]{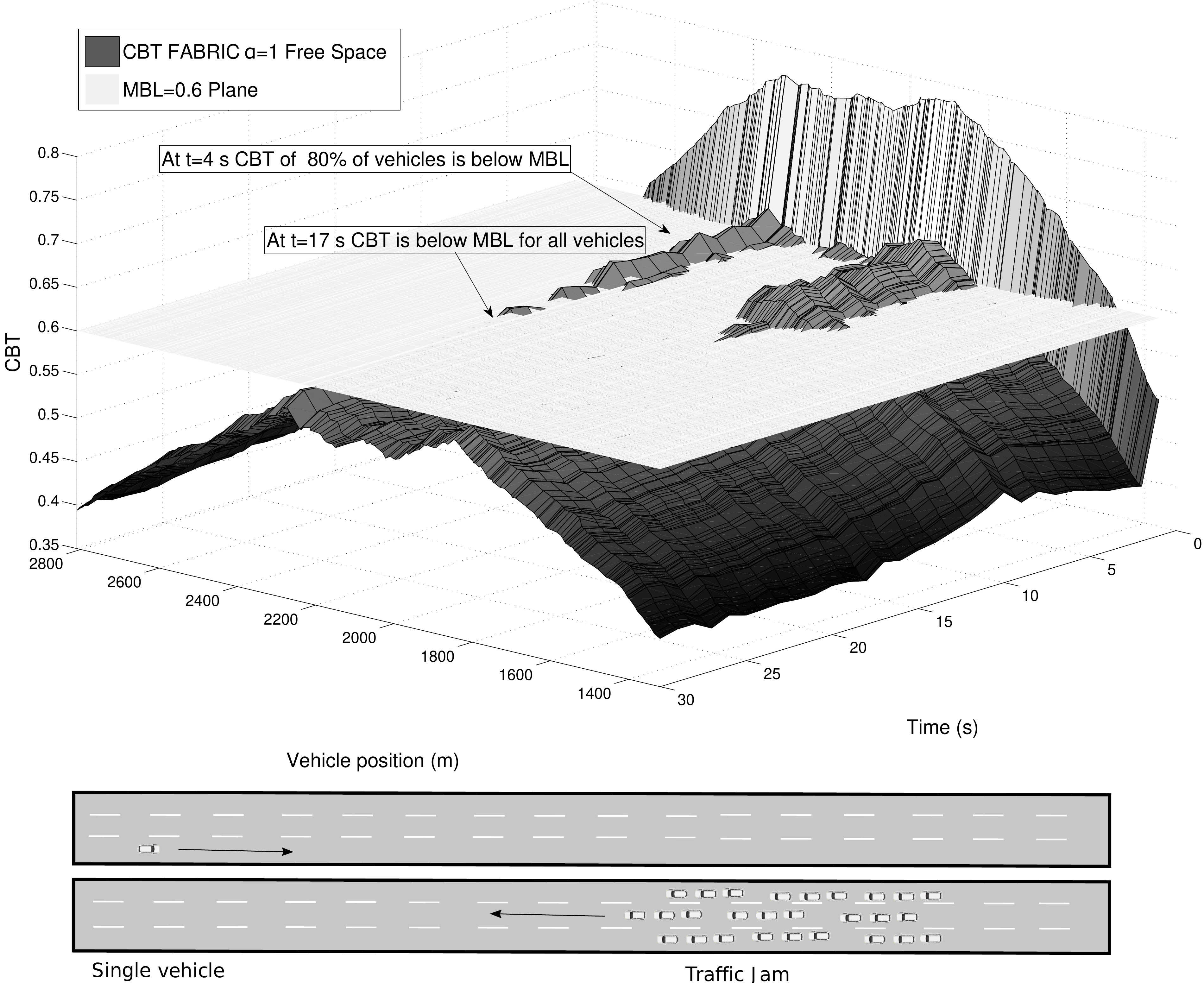}%
	\label{cbtonramp}
}
	\caption{Vehicle approaching a cluster of 200 motionless nodes. FABRIC asynchronous and  LIMERIC+PULSAR.}
\label{Ramp}

\end{figure*}
In this section we investigate the FABRIC performance in dynamic configurations, where vehicles move. We look mainly at the time evolution of the selected beaconing rate and CBT.
Our goal is to test the ability of FABRIC to perform smooth transitions from low to high congestion situations and whether it results in oscillations.

\subsection{ Single vehicle and traffic jam}  
The first scenario involves (i) a cluster of statically positioned vehicles along a 1500 m road segment, using the same random Poisson positioning of 
 Section \ref{mhsimscenario}, and (ii) a single vehicle approaching the cluster, starting 1320 m away from the last cluster car, and moving at a constant speed of 32 m/s until it passes the cluster. This configuration can model different real scenarios such a  highway with a traffic jam in one direction and a single vehicle moving in the opposite direction.
The goal of this configuration is to show the dynamics of FABRIC in an extreme case where a vehicle switches from no or very few neighbors to a congested state, and back again. Let us note that, with the selected parameters shown in Table \ref{tcbt}, 
a channel can accommodate approximately 78 vehicles in range at the maximum rate. The case of a congested cluster approaching another one is considered in the next configuration.
Finally, let us remark again that MAC collisions, hidden node interference and propagation errors are present in these scenarios. 
Since we are interested in the time evolution of the variables we plot only the results of one replication (all of them show a similar evolution). 

\textit{Beaconing rates}. In Fig. \ref{onramp} we show the time evolution of the beaconing rate of the moving vehicle for FABRIC and LIMERIC+PULSAR. 
Beaconing rates of the cluster vehicles are not shown since the effect of the single moving vehicle in their rate is negligible.
With both fairness notions  FABRIC can keep the single vehicle at the maximum rate until the vehicle is in range of at least 
78 neighbors, at $t=40.8$ s. Afterwards, it reduces its rate according to the state of the channel in practically the same way as its neighbors (compare with Fig \ref{brmulti}). This happens both in 
free-space and Nakagami-m configurations. In the latter, variations are smoother in the middle, 
showing that some earlier pike effects of packet losses caused by  fading are compensated in the presence of a large number of neighbors (e.g. in congested areas),
 and at the same time convergence is faster since the feedback (number of received prices) is higher. 
The effective beaconing rate $\hat r_v(250)$ shows the same trend that as in the previous section, with an average drop of 20\% at 250 m.
In its turn, the moving LIMERIC vehicle reduces the rate earlier than necessary and recovers it later, even with free space propagation.
Actually it does not recover the maximum rate until it is completely out of range at $t=105.8$ s. It also shows a characteristic oscillatory behavior of LIMERIC+PULSAR \cite{Pulsar}. 

\textit{CBT}. Finally, Fig. \ref{cbtonramp} shows the CBT of all the vehicles for FABRIC with $\alpha=1$ . Interestingly, FABRIC quickly moves rates out of congestion at the beginning.
 Moreover, these results which are the time 
evolution of those in Fig. \ref{brmulti} bottom confirm 
that it is  not necessary to achieve the optimal allocation to meet the MBL constraint, showing that after only a few steps at t=4 s, 80\% of the vehicles measure a load below the MBL. In fact, as  $\alpha$ increases both 
beaconing rate and CBT (not shown in the figure) show a steeper reaction to congestion, that is, in a congested state vehicles quickly and abruptly reduce their rates and then increase them until convergence is
achieved. 

\subsection{ Bridge over highway}
\begin{figure*}[!t]
	\centering
\subfloat[Beaconing rates versus time for the cluster of 100 vehicles crossing the bridge at 32 m/s. Top: FABRIC. Bottom: LIMERIC+PULSAR ] {		
\includegraphics[width=0.96\columnwidth]{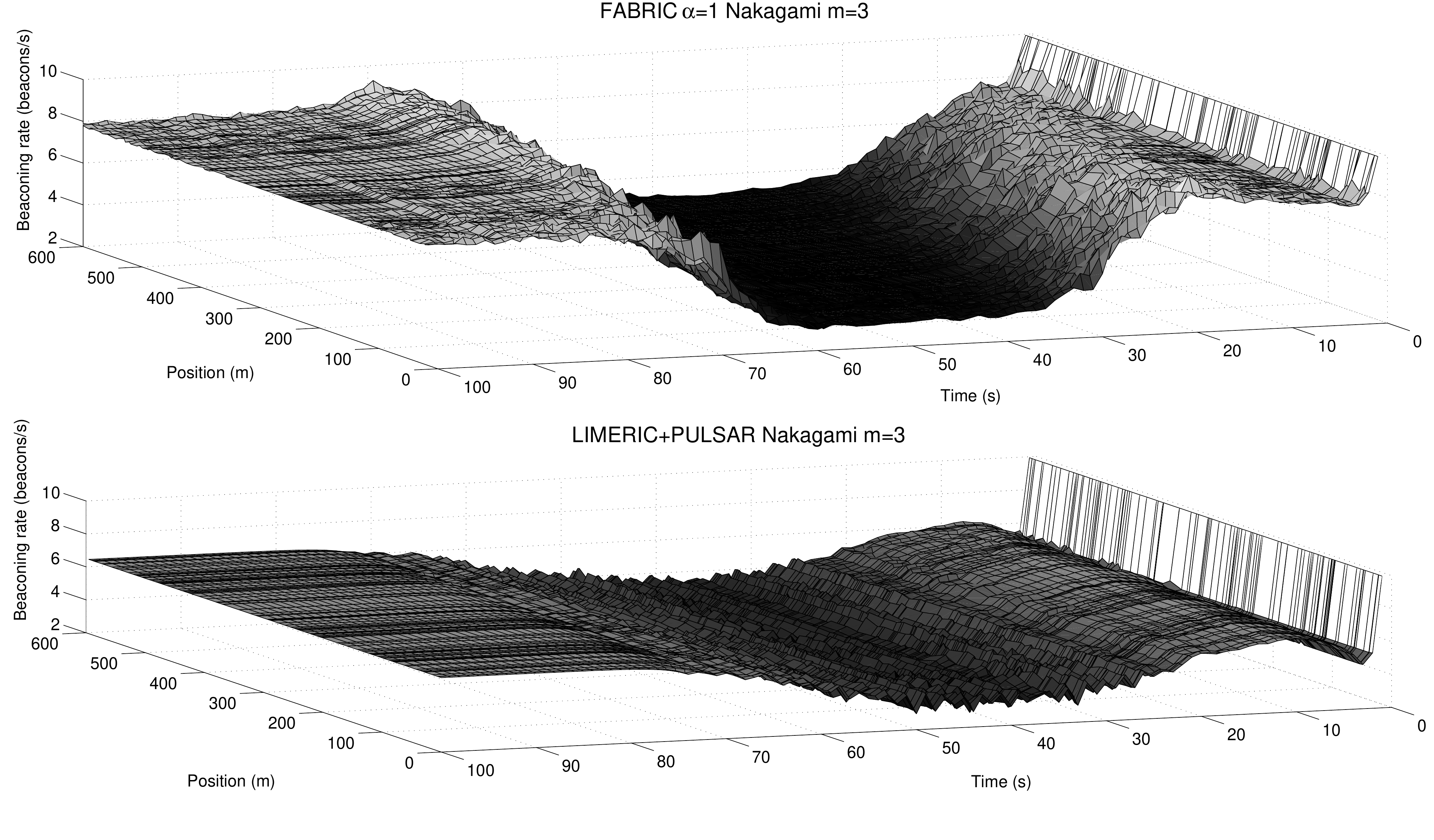}%
	\label{bridgeUpRate}
}
\hfill 
\subfloat[CBT versus time for the cluster of 100 vehicles crossing the bridge. Top: FABRIC. Bottom: LIMERIC+PULSAR ]{
		\includegraphics[width=0.96\columnwidth]{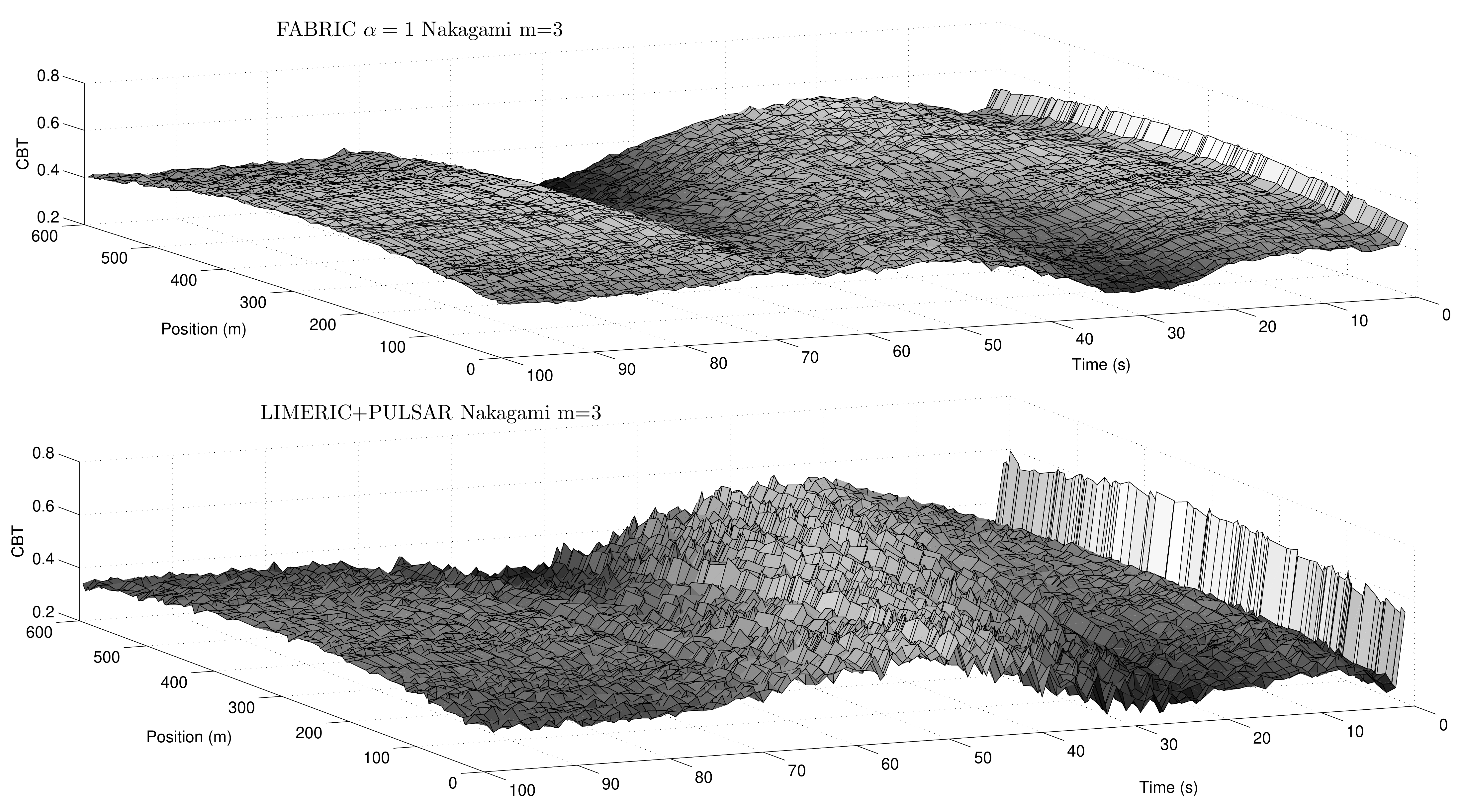}%
	\label{bridgeUpCBT}
}
	\caption{Beaconing rates and CBT versus time for a cluster of 100 nodes  crossing a bridge over a highway at 32 m/s.   Nakagami m=3 propagation. 
FABRIC asynchronous  and  LIMERIC+PULSAR.}
\label{FSBridge}
\end{figure*}
\begin{figure*}[!t]
	\centering
\subfloat[Beaconing rates versus time for the cluster of 200 static vehicles on the highway. Top: FABRIC. Bottom: LIMERIC+PULSAR ]{
		\includegraphics[width=0.96\columnwidth]{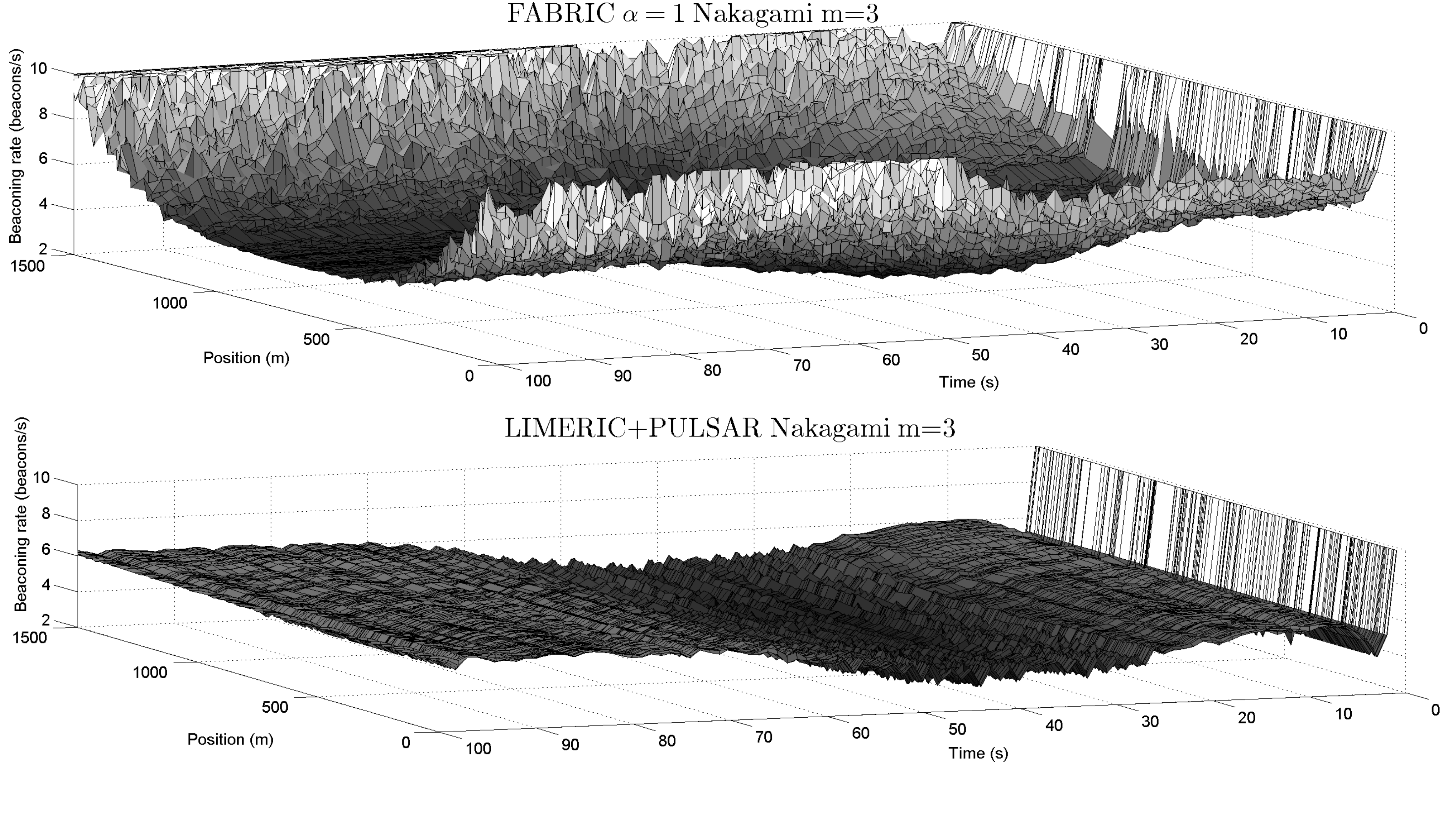}%
	\label{bridgeDownRate}
}
\hfill 
\subfloat[CBT versus time for the cluster of 200 static vehicles on the highway. Top: FABRIC. Bottom: LIMERIC+PULSAR]{
		\includegraphics[width=0.96\columnwidth]{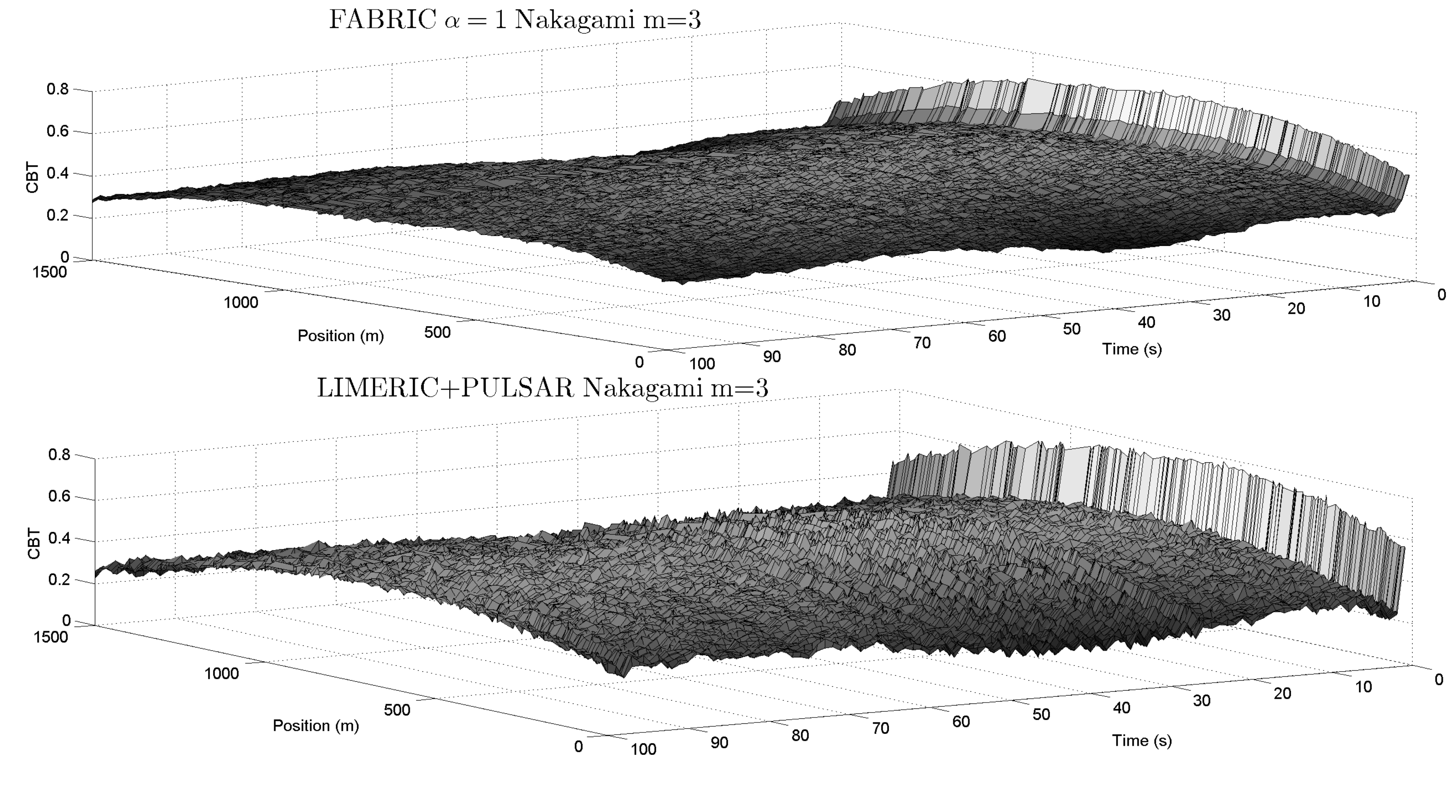}%
	\label{bridgeDownCBT}
}
	\caption{Beaconing rates and CBT versus time for a cluster of 200 static vehicles on the highway.   
  Nakagami m=3 propagation. 
FABRIC asynchronous  and  LIMERIC+PULSAR.}
\label{FSBridgeDown}
\end{figure*}
Finally, we examine a extreme scenario, where a static cluster of approximately 200 vehicles is set along a 1500 m highway segment oriented north-south ($y$ axis), and it is 
crossed at the middle position by another highway west-east oriented ($x$ axis). 
A bridge is situated at the crossing, so that the west-east highway passes over the north-south highway. 
A moving cluster of 100 vehicles, stretching over 600 m,  moves from west to east at a constant speed of 32 m/s, starting 1500 m away from the bridge. The moving cluster approaches the static cluster, crosses the bridge, and 
moves away. For both clusters, the initial position and propagation configurations have been set as in the previous scenarios: Poisson and free space and Nakagami-m and again the transmission range results in hidden nodes. Due to lack of space, only the Nakagami-m results are shown, yielding to the same conclusions as the ones for free-space.

In Fig. \ref{FSBridge} and \ref{FSBridgeDown} we show the time evolution of the  beaconing rates  and CBT for all the vehicles in both clusters. In both cases, FABRIC reduces  
the rate of the vehicles as the transmission ranges overlap, but never below 3 beacons/s in any cluster. 
Fig. \ref{bridgeUpRate} illustrates the evolution of the rate allocations in the moving cluster. With FABRIC, vehicles at the rear increase at the beginning their rates, 
as the front ones reduce theirs when they are entering the range of the highway vehicles, and conversely as they move away from the bridge.  The time evolution of the rates of the static vehicles plotted in Fig. \ref{bridgeDownRate} shows how the 
central vehicles reduce their rates when the cluster passes and recover them later. The rate variations are small  at the center and  higher at the vehicles on the edges, since the feedback from their neighbors is weaker. 
As shown in Fig. \ref{bridgeUpCBT} and Fig. \ref{bridgeDownCBT}, FABRIC also succeeds in keeping the channel at an allowable  utilization (approx. 50\%). Results for LIMERIC+PULSAR show again 
that it tends to assign the same rate to all the vehicles and exhibits an oscillatory behavior in 
both rates and CBT when both clusters are in range of each other.
\subsection{Traffic queue }
\label{Queue}
\begin{figure}[!t]
	\centering
		\includegraphics[width=\columnwidth]{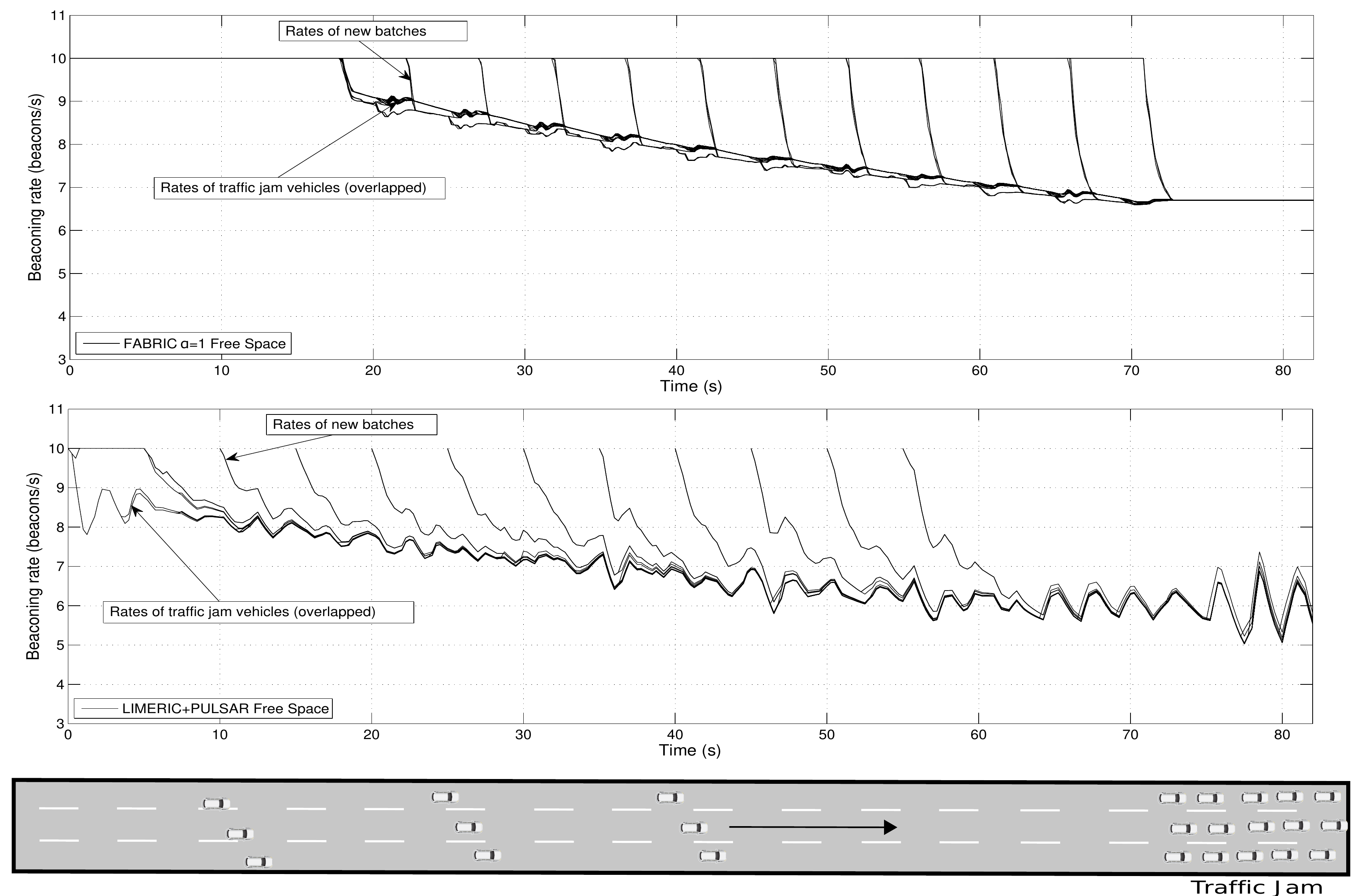}
	\caption{Vehicles arrive at a traffic jam in batches, building up a queue. Beaconing rate time evolution for FABRIC (top) and LIMERIC+PULSAR (bottom). A schematic drawing is shown below. }
	\label{queue-fig}
\end{figure}

In the previous subsections we have shown worst case scenarios, where either a large group of vehicles start simultaneously from an highly congested state or merge together. In this last scenario we 
look at a more likely situation, where the congestion is building up progressively. Our goal is to examine the behavior and convergence time of the algorithms. 
In Fig. \ref{queue-fig} we show the time evolution of the beaconing rates in an scenario where there is a static cluster of 76 vehicles all in range of each other. There is no congestion, since 
it is below the  78 vehicles transmitting at maximum rate that allows the MBL. Batches of 3 new vehicles are added at the origin 700 m away and move at 32 m/s until they reach the end of the queue and then stop. 
A new batch is created every 5 s. This scenario may model an on-ramp where vehicles join a  jammed highway.

As can be seen, with FABRIC both vehicles in the jam and  vehicles in the batches keep the maximum rate until they actually contribute to congestion, when they are in range of each other after 18 s.
 On the contrary, with LIMERIC+PULSAR 
vehicles again  start reducing the rate earlier because of the two-hop congested state they are receiving as feedback from the preceding batch. 
Regarding convergence time of the vehicles in the batches,  defined as the time interval elapsed since a vehicle start to reduce the 
rate until it has reached the value of the queue, FABRIC shows an average of 1.73 s compared to 5.45 s of LIMERIC+PULSAR.



\section{Conclusion and future work}
\label{Conclusions}

In this paper we model for the first time, to the best of our knowledge, the problem of beaconing rate control in vehicular networks  as a NUM rate allocation problem. 
This modeling opens the door to formally define and apply fairness notions to beaconing rate allocations in vehicles. 
In addition, it provides a mathematical framework to develop decentralized and  simple algorithms with proved convergence guarantees to a \textit{fair} allocation solution. 
In this respect, we have presented a family of algorithms based on the gradient optimization of the dual of the rate allocation problem. Within this family,  
 we have proposed the Fair Adaptive Beaconing Rate for Intervehicular Communications (FABRIC) algorithm. FABRIC, is a decentralized rate allocation algorithm with 
 theoretical and empirical convergence properties, which requires limited  signaling overhead between vehicles. 

We have validated FABRIC by exhaustive simulations in both static and dynamic scenarios, for different position distributions and propagation models. 
Results show that FABRIC effectively generates fair beaconing rates allocations. Moreover, only in  a few steps, the algorithm is able to move the rates out of the congestion state 
and close to the optimal allocation. 
Simulations also confirm  that the algorithm is robust against packet losses due to collisions or fading.
Our results have been compared with LIMERIC+PULSAR, a relevant rate allocation scheme in vehicular networks. 

There are still a number of practical considerations and implementation alternatives that can be evaluated in order to tune the algorithm. First, the $\beta$ parameter controls the 
convergence speed and the amplitude of fluctuations and there is a wide range of possible values meeting the convergence condition to test. 
Second,  filtering of unreliable links 
 may provide a more accurate   measurement of congestion (gradient computation)  in fading scenarios. Even  the use of alternative congestion measurements such as the measured CBT can be tested. 
We intend to carry out an extensive evaluation of these matters in a future work.

Additionally, from a more general perspective, we have shown how 
different values of the fairness parameter $\alpha$ result in different allocations,  which may be more adequate depending on the intended application or scenario. 
As we discussed with a particular example, a too basic fairness notion may directly influence the safety of the vehicles. Therefore, it is necessary to study  which is the appropriate notion of fairness in vehicular networks and 
whether different scenarios require different notions of fairness. This is  an open question left as future work but a positive answer implies that 
it is also necessary a mechanism to dynamically control fairness. 
In this sense, one of the key advantages of FABRIC and our approach is that the fairness allocations can be controlled with this single parameter. Moreover, this approach allows to use different values 
for each vehicle or even to use totally different utility functions, which can be both dynamically changed. And vehicles do not need to know the functions or values used by other vehicles.
Therefore, in addition to the practical utility of our proposal, in our opinion, one of the main contributions of this paper is the establishment of the NUM model as an effective and rich  framework for developing 
beaconing rate control schemes in vehicular networks. 
Consequently, as a future work, we intend to further explore variations of the discussed problem in the context of vehicular networks. In particular, a comparative  application and evaluation of alternative 
fairness notions  and the introduction of heterogeneous utility functions and constraints in the problem. 

Finally, our model also provide support for the quality of service needs of the applications, which usually require to control additional variables such as transmit power. In fact, many recent proposal let 
the  application freely set  minimum values for one or severals variables and then apply a distributed control for the rest of them 
over the remaining capacity, which  might results in violations of the MBL.
It is interesting to have a more integrated approach and so we are working on a reformulation of the problem to jointly control power and beaconing rate.

\ifCLASSOPTIONcompsoc
  \section*{Acknowledgments}
\else
  \section*{Acknowledgment}
\fi

This research has been supported by the MINECO/FEDER project grant TEC2013-47016-C2-2-R (COINS) and 
 partially supported by the Spanish project grants TEC2014- 53071-C3-1-P (ONOFRE) and TEC2015-71932-REDT (ElasticNetworks)

\ifCLASSOPTIONcaptionsoff
  \newpage
\fi




\begin{thebibliography}{1}

\bibitem{VanetBook} H. Hartenstein and K. P. Laberteaux, 
``VANET. Vehicular Applications and Inter-Networking Technologies'', Wiley, 2010. 

\bibitem{EtsiPhy} 
ETSI EN 302 663, ``Intelligent Transport Systems (ITS); Access layer specification for Intelligent Transport Systems operating in the 5 GHz frequency band'', 
V0.1.3, 2012

\bibitem{802.11p}
IEEE Std 802.11p-2010, ``IEEE Standard for Information technology-- Local and metropolitan area networks-- Specific requirements-- Part 11: Wireless LAN Medium Access Control (MAC) and Physical Layer (PHY) Specifications Amendment 6: Wireless Access in Vehicular Environments,''  (Amendment to IEEE Std 802.11-2007 as amended by IEEE Std 802.11k-2008, IEEE Std 802.11r-2008, IEEE Std 802.11y-2008, IEEE Std 802.11n-2009, and IEEE Std 802.11w-2009) ,  pp. 1-51,  2010.

\bibitem{Survey}
G. Karagiannis, O. Altintas, E. Ekici, G. Heijenk, B. Jarupan, K. Lin, and T. Weil,
``Vehicular Networking: A Survey and Tutorial on Requirements, Architectures, Challenges, Standards and Solutions,''
{\em IEEE Communications Surveys \& Tutorials}, vol. 13, no. 4, pp. 584-616, 2011.

\bibitem{EtsiCAM} 
ETSI EN 302 637-2, ``Intelligent Transportation Systems (ITS); Vehicular Communications; Basic Set of Applications; Part 2: Specification 
of Cooperative Awareness Basic Service'', 2013.

\bibitem{Awareness}
M. Sepulcre, J.  Mittag, P.  Santi, H. Hartenstein, and J.  Gozalvez,  ``Congestion and Awareness Control in Cooperative Vehicular Systems,''
{\em Proceedings of the IEEE}, vol. 99, no. 7, pp. 1260-1279,  2011.

\bibitem{ChiangFair} 
K, Tian Lan,  M. Chiang, A.  Sabharwal, ``An Axiomatic Theory of Fairness in Network Resource Allocation,''
{\em Proceedings IEEE INFOCOM},  pp.1--9,  2010.

\bibitem{Limeric}
J. B. Kenney, G. Bansal and C. E. Rohrs, ``LIMERIC: A Linear Adaptive Message Rate Algorithm for DSRC Congestion Control,''
{\em  IEEE Transactions on Vehicular Technology}, vol. 62, no 9, pp. 4182--4197, 2013.

\bibitem{Pulsar}
T. Tielert, D. Jiang, Q. Chen, L. Delgrossi and H. Hartenstein, ``Design Methodology and Evaluation of Rate Adaptation Based Congestion Control for Vehicle Safety Communications,''
{\em Vehicular Networking Conference (VNC), 2011 IEEE},  pp. 116-123,  2011.

\bibitem{Fallah}
Y. P. Fallah, C. L. Huang, R. Sengupta aand H. Krishnan, ``Analysis of Information Dissemination in Vehicular Ad-Hoc Networks With Application to Cooperative Vehicle Safety Systems,''
{\em  IEEE Transactions on Vehicular Technology}, vol. 60, no 1, pp. 233--247, 2011.

\bibitem{Intern}
M. Sepulcre, J. Gozalvez, O. Altintas and H. Kremo , ``Adaptive Beaconing for Congestion and Awareness Control in Vehicular Networks,''
{\em Vehicular Networking Conference (VNC), 2014, IEEE},  pp. 81-88,  2014.

\bibitem{Tielert}
T. Tielert, D. Jiang,   H. Hartenstein and L. Delgrossi , ``Joint Power/Rate Congestion Control Optimizing Packet Reception in Vehicle Safety Communications,''
{\em Proceedings of the tenth ACM international workshop on Vehicular inter-networking, systems, and applications},  pp. 51-60,  2013.



\bibitem{Kelly99} Kelly, Frank, ``Charging and rate control for elastic traffic.'' European transactions on Telecommunications 8.1 (1997): 33-37.

\bibitem{MoWal}
J. Mo and J. Walrand, ``Fair end-to-end window-based congestion control,''
{\em  IEEE/ACM Transactions on Networking}, vol. 8, no. 5, pp. 556--567,  2000.
\bibitem{EtsiDCC} 
ETSI TS 102 687, ``Intelligent Transport Systems (ITS); Decentralized Congestion Control Mechanisms for Intelligent
Transport Systems operating in the 5 GHz range; Access Layer part'',  2011.

\bibitem{Kim}
B. Kim, I. Kang, H. Kim, ``Resolving the Unfairness of Distributed Rate Control in the IEEE WAVE Safety Messaging,''
{\em IEEE Transactions on Vehicular Technology}, vol. 63, no. 5, pp. 2284-2297, 2014.

\bibitem{SBCC}
E. Egea-Lopez, J. J.  Alcaraz, J. Vales-Alonso, A Festag and J. Garcia-Haro, ``Statistical Beaconing Congestion Control for Vehicular Networks,''
{\em IEEE Transactions on Vehicular Technology},   vol. 62, no. 9, pp. 4162--4181,  2013.

\bibitem{EMBARC}
G. Bansal, H. Lu, J. B. Kenney, C. Poellabauer, ``EMBARC: error model based adaptive rate control for vehicle-to-vehicle communications,''
{\em  Proceeding of the tenth ACM international workshop on Vehicular inter-networking, systems, and applications}, pp. 41-50, 2013.

\bibitem{Boyd}
S. Boyd and L. Vandenberghe,  
{\em Convex Optimization}. Cambridge University Press, Cambridge, UK, 2004. 

\bibitem{Bertsekas}
D. P. Bertsekas and J. N. Tsitsiklis, 
{\em Parallel and Distributed Computation: Numerical Methods}. Prentice-Hall, Inc., Upper Saddle River, NJ, 1989. 


\bibitem{Chiang}
M. Chiang, S. H.  Low,  A. R.  Calderbank, and J. C.  Doyle, ``Layering as optimization decomposition: A mathematical theory of network architectures,''
{\em  Proceedings of the IEEE}, vol 95, no 1, pp.  255--312.

\bibitem{Sri2014}
R. Srikant and L. Ying, {\em Communication networks. An optimization, control and stochastic networks perspective}. Cambridge University Press, 2014. 

\bibitem{Low}
S. Low and D. Lapsley, ``Optimization Flow Control--I: Basic Algorithm and Convergence,''
{\em  IEEE/ACM Transactions on Networking (TON)}, vol. 7, no. 6, pp. 861-874, 1999.


\bibitem{Subramanian}
S. Subramanian, M. Werner, S. Liu, J. Jose, R. Lupoaie and X. Wu, ``Congestion control for vehicular safety: synchronous and asynchronous MAC algorithms,''
{\em Proceedings of the ninth ACM international workshop on Vehicular inter-networking, systems, and applications}, pp. 41-50, 2012. 

\bibitem{ChanJSAC}
L. Cheng, B. E. Henty, D. D. Stancil, F. Bai, and P. Mudalige, ``Mobile Vehicle-to-Vehicle Narrow-Band Channel Measurement and Characterization of the 5.9 
GHz Dedicated Short Range Communication (DSRC) Frequency Band,'' 
{\em IEEE Journal on Selected Areas in Communications}, vol. 25, no. 8, pp. 1501--1516, 2007.

\bibitem{Bertsekas2}
D. P. Bertsekas, ``Projected Newton methods for optimization problems with simple constraints,'' 
{\em SIAM Journal of Control and Optimization}, vol. 20, no. 2, 1982.


\bibitem{itsworkshop}
J. Kenney, D. Jiang, G. Bansal, T. Tielert, ``Controlling Channel Congestion using CAM Message Generation Rate'', 
{\em 5th ETSI ITS Workshop}, 2013. 

\bibitem{omnetpp}
OMNeT++ simulation library, avaliable at http://www.omnetpp.org
%
%
%
\bibitem{JOM}
JOM (Java Optimization Modeler), avaliable at http://www.net2plan.com/jom/

\end{thebibliography}
%

%

\begin{IEEEbiography}[{\includegraphics[width=1in,height=1.25in,clip,keepaspectratio]{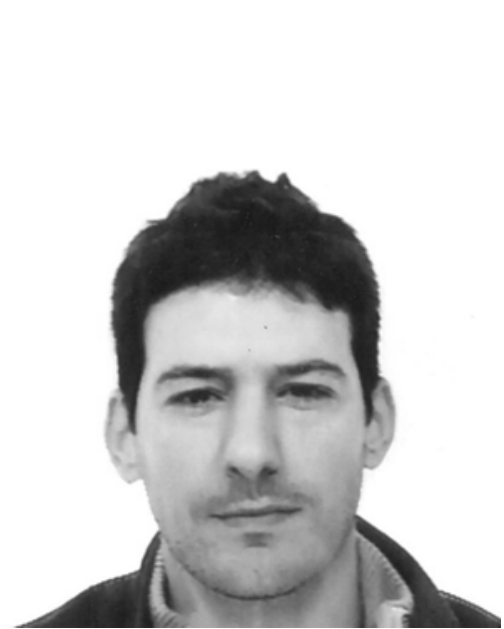}}]{Esteban Egea-Lopez}
received the Telecommunications Engineering degree in 2000, from the Universidad Politecnica de Valencia (UPV), Spain, the Master Degree in Electronics in 2001, 
from the University of Gavle, Sweden, and Ph.D. in Telecommunications in 2006 from the Universidad Politecnica de Cartagena (UPCT).

He is an associate professor with the Department of Information Technologies and Communications at the Universidad Politecnica de Cartagena (UPCT). His research interest is focused on vehicular networks 
and MAC protocols.
\end{IEEEbiography}

\begin{IEEEbiography}[{\includegraphics[width=1in,height=1.25in,clip,keepaspectratio]{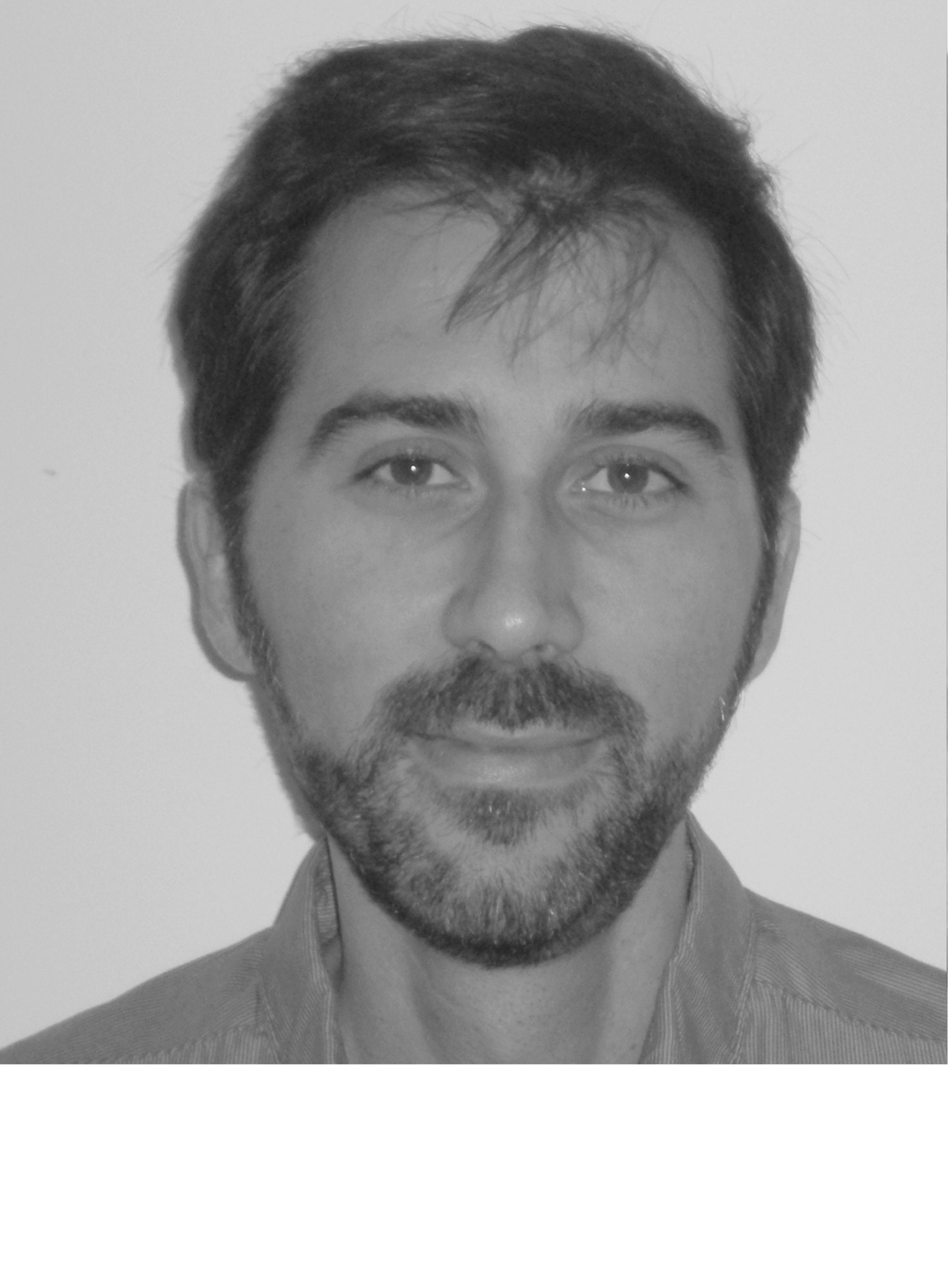}}]{Pablo Pavon-Mari\~no}
 received the M.Sc. degree in Telecommunication Engineering in 1999 from the University of Vigo, Spain. In 2000, he joined the Universidad Politécnica de Cartagena, Spain, where he is Associate Professor in the Department of Information Technologies and Communications. He received the Ph.D. degree from this University in 2004, and a M.Sc. degree in Mathematics in 2010. His research interests include 
planning and optimization of communication networks.
\end{IEEEbiography}







\end{document}